\documentclass[floatfix,groupedaddress,superscriptaddress,aps, amsmath,amssymb, pra,longbibliography, twocolumn,a4]{revtex4-1}

\usepackage{graphicx,float}
\usepackage{subfigure}
\usepackage{dcolumn}
\usepackage{bm}
\usepackage{mathrsfs}
\usepackage{txfonts}
\usepackage{CJK}
\usepackage{epsfig}
\usepackage{lipsum}
\usepackage{color}
\usepackage{placeins}
\usepackage[colorlinks]{hyperref}

\makeatletter

\newcommand{\Rmnum}[1]{\expandafter\@slowromancap\romannumeral #1@}
\makeatother

\hypersetup{    plainpages=true,
    breaklinks=true,    hypertexnames=false,    pageanchor=true,
    colorlinks=true,
    linkcolor={blue},
    citecolor={blue},
    urlcolor={blue},
    anchorcolor={black}
}

\begin{document}

\title{Chaotic synchronization of two optical cavity modes in optomechanical systems}

\author{Nan Yang}\email{nan.yang@riken.jp}
\affiliation{CEMS, RIKEN, Saitama, 351-0198,
Japan}\affiliation{National Laboratory of Solid State
Microstructures, College of Engineering and Applied Sciences,
Nanjing University, Nanjing, 210093, China}

\author{Adam Miranowicz}
\affiliation{CEMS, RIKEN, Saitama, 351-0198, Japan}
\affiliation{Faculty of Physics, Adam Mickiewicz University,
61-614 Poznan, Poland}

\author{Yong-Chun Liu}
\affiliation{State Key Laboratory of Low Dimensional Quantum
Physics and Department of Physics, Tsinghua University, Beijing,
100084, China}

\author{Keyu Xia}\email{keyu.xia@nju.edu.cn}
\affiliation{National Laboratory of Solid State Microstructures,
College of Engineering and Applied Sciences, Nanjing University,
Nanjing, 210093, China} \affiliation{Collaborative Innovation
Center of Advanced Microstructures, Nanjing 210093, China}\affiliation{CEMS, RIKEN, Saitama, 351-0198, Japan}

\author{Franco Nori}
\affiliation{CEMS, RIKEN, Saitama, 351-0198, Japan}
\affiliation{Physics Department, The University of Michigan, Ann
Arbor, Michigan 48109-1040, USA}
\date{\today}

\begin{abstract}
The synchronization of the motion of microresonators has attracted
considerable attention. Here we present theoretical
methods to synchronize the chaotic motion of two optical cavity
modes in an optomechanical system, in which one of the optical
modes is strongly driven into chaotic motion and is coupled to
another weakly-driven optical mode mediated by a mechanical
resonator. In these optomechanical systems, we can obtain both complete
and phase synchronization of the optical cavity modes in chaotic
motion, starting from different initial states. We find that complete synchronization of chaos can be achieved in
two identical cavity modes. In the strong-coupling
small-detuning regime, we also {produce} phase synchronization of chaos
between two nonidentical cavity modes.
\end{abstract}

\maketitle

\section{\label{sec:level1}Introduction\protect\\ }

The synchronization of oscillators is a universal concept in
nonlinear sciences~\cite{review1,review2}. It has been observed in
both nature~\cite{review2} and social activities~\cite{review1,review2,social},
and also promises important applications in
engineering~\cite{review1,review2,signalprocessing,communication,clock1}. Since
its discovery in pendulum systems by Huygens in the 17th
century~\cite{Huygens}, synchronization has been observed in
various fields including bursting neurons~\cite{neuro},
fireflies~\cite{glowworms}, and chemical
reactions~\cite{chemistry}. Although these systems operate in very
different size scales, the mechanism behind synchronization can be
understood as follows: oscillators under weak interaction adjust
their rhythms to keep their motions consistent. The
synchronization of oscillators has been studied in relation to
information processing~\cite{signalprocessing},
{communications}~\cite{communication}, and high-precision
clocks~\cite{clock1}.

Optomechanical resonators~\cite{OM1,OM3,OM4,OM5,OM6,
OM8,OM9,OM10,Xiao2017,Sciamanna2016,OM13,OM14,OM15,OM16} with high-quality factors and
strong nonlinearities have attracted {considerable} attention in various fields
due to their promising applications. The synchronization of
optomechanical systems is an important topic in
optomechanics~\cite{experiments1,experiments2,experiments4,experiments5,
optosyn1,optosyn2,optosyn3,optosyn4,experiments6,experiments7}.
However, the majority of previous works concentrate on the
synchronization of \emph{periodic} oscillations. The
\emph{chaotic} synchronization~\cite{chaoticsyn1} of an
optomechanical system is very challenging because chaotic signals
are extremely sensitive to initial conditions. It is still an open
question whether microscopic optomechanical systems with chaotic
motion can be synchronized. Optomechanical systems with strong
nonlinear light-matter interactions can support quite
different types of motion, i.e., periodic~\cite{OM3},
quasi-periodic~\cite{OM13}, and chaotic~\cite{Xiao2017,Sciamanna2016,OM13,OM14,OM15,OM16}.
Thus, the study of chaotic synchronization of optomechanical
systems may provide an answer to this question.

In this paper, we study both complete and phase synchronization of
two optical cavity modes in an optomechanical system with chaotic
dynamics rather than with periodic motion. We consider an optical
cavity mode strongly driven to a chaotic state. This brings other
weakly-driven optical-cavity modes into chaos mediated by a
mechanical motion~\cite{OM13}. Thus, it is found that complete
synchronization is achievable in two identical weakly-driven
cavity modes, and phase synchronization can be realized in the
strongly- and weakly-driven cavity modes, although these two optical modes are initially in different states.

The active-passive decomposition (APD) model~\cite{APD1,APD2} is
widely used to describe systems in complete synchronization. In
the APD model, different subsystems under a common driving force
can achieve complete synchronization regardless of their initial
conditions. This APD model can also describe our chaotic system.
The complete synchronization studied here (as shown in
Fig.~\ref{fig01}) involves three cavity modes coupled to a common
mechanical oscillator in an optomechanical system. One of the
cavity modes, $\hat{a}_s$, is strongly driven into
chaotic motion. Mediated by the mechanical oscillation, it also
drives the other two weakly-driven optical cavity modes,
$\hat{a}_1$ and $\hat{a}_2$, into chaotic motion. We show that
this chaotic motion of the modes $\hat{a}_1$ and $\hat{a}_2$ can
be completely synchronized.
%
However, this complete synchronization can only be realized in
identical chaotic systems. A small mismatch in design of two
synchronized subsystems can destroy their complete synchronized
behavior.

In contrast to complete synchronization, phase synchronization can
be realized between two nonidentical chaotic systems through
interacting with a common mechanical motion (see
Fig.~\ref{fig02}). To achieve such phase synchronization, one
optical cavity mode ($\hat{a}_s$) in the optomechanical system is
strongly driven by an external classical field, while the other
mode ($\hat{a}_w$) is weakly driven. Phase synchronization of
the optical modes can be achieved when the cavity-driving detuning
is much smaller than the optomechanical coupling. We find that the
temporal phases of the two optical modes with chaotic motion
mainly depend on the mechanical displacement, which is governed by
the cavity mode $\hat{a}_s$. Thus, although the two optical modes
are in chaotic motion, their unwrapped phases can be locked to
each other at a fixed ratio.
In the following, we investigate these two kinds of
synchronization in two different models.

We propose two setups (A and B) for either complete synchronization or phase synchronization of the optical modes in an optomechanical system. Both setups A and B share {a} common configuration: the strongly-driven cavity mode $\hat{a}_s$ dominates the {motion} of the weakly-driven cavity modes. In setup A, the strongly- and weakly-driven cavity modes are coupled via a mechanical mode, while they are coupled via two mechanical resonators in setup B. In comparison with the setup A, a strong coupling between two mechanical resonators is required in setup B.

This paper is organized as follows: In Secs.~\Rmnum{2}
and~\Rmnum{3}, {we present the corresponding
setups for both complete and phase synchronization in an
optomechanical system.} The numerical results for these two types
of synchronization are shown and compared in Sec.~\Rmnum{4}. In
Sec.~\Rmnum{5}, we summarize our work and discuss some potential
applications.

\section{Complete synchronization}

\begin{center}
\begin{figure*}[t]
\centering
\includegraphics[width=6in]{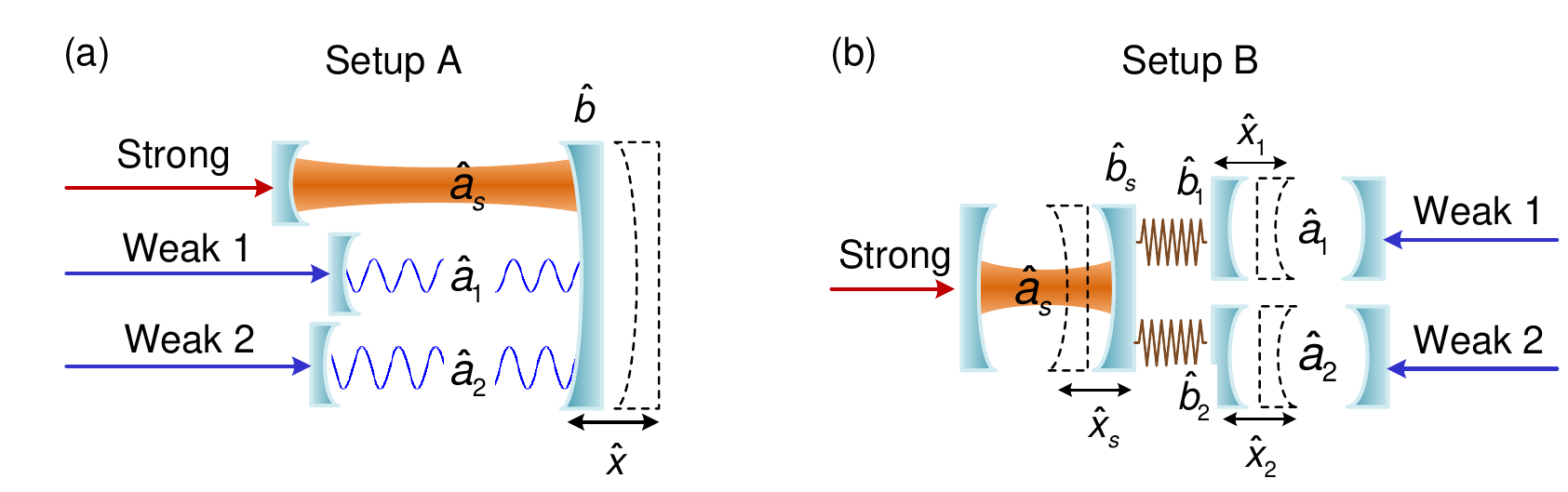}
\caption{(Color online) Schematic diagrams of two optomechanical
models for complete synchronization. (a) Setup~A  includes a
strongly-driven cavity mode $\hat{a}_s$, two weakly-driven cavity modes $\hat{a}_j$, and a
mechanical mode $b$. The strongly- and weakly-driven cavity modes are
coupled via the mechanical mode {with coupling strengths $g_s$ and $g_j$}, respectively. (b) {Setup~B consists of a
strongly-driven cavity mode $\hat{a}_s$ and two weakly-driven cavity modes $\hat{a}_j$,
where the latter are coupled to $\hat{a}_s$ via the mechanical modes $b_j$ and $b_s$, which are additional coupled to each other with the spring
coefficients $k_j$}.} \label{fig01}
\end{figure*}
\end{center}

In this section, we focus on the complete synchronization of an
optomechanical system. In general terms, complete synchronization
refers to the identity among the phase-space orbits of chaotic
systems. Let us consider two chaotic systems
\begin{equation}
\dot{\textbf{y}}_1=\textbf{f}(\textbf{y}_1), \qquad
\dot{\textbf{y}}_2=\textbf{f}(\textbf{y}_2),
\end{equation}
where $\textbf{y}_1$ and $\textbf{y}_2$ are $N$-dimensional
variables governed by the function $\textbf{f}: R^N \rightarrow
R^N$. We define the difference between the phase-space orbits of
two chaotic systems as the synchronization error $\textbf{e}(t)$, {where}
$\textbf{e}(t)=\textbf{y}_1(t)-\textbf{y}_2(t)$. Two chaotic
systems are called completely synchronized if and only if their
synchronization error $\textbf{e}(t)$ vanishes in the evolution {long}
time limit~\cite{CS1}, i.e.,
\begin{equation}
\lim_{t\rightarrow\infty}\textbf{e}(t)=\lim_{t\rightarrow\infty}\|\textbf{y}_1(t)-\textbf{y}_2(t)\|=0.
\end{equation}


The drive-response  model~\cite{CS1} and the active-passive
decomposition (APD) model~\cite{APD1,APD2} are two widely used
methods for characterizing the complete synchronization of chaotic
systems. In the former model, the drive and response systems,
which are to be synchronized, are in the unidirectional-coupling
regime. It is required that the response system can be decomposed
into a stable subsystem and an unstable one. By controlling the
motion of the unstable subsystem, the {driving part} can force the
phase-space orbit of the response part to reach a synchronized
state. However, the drive-response model can only be applied to
decomposable chaotic systems. This seriously restricts its
applications in engineering. The APD {model}{,} as an advanced
version of the drive-response {model}{,} provides a more general way
to study complete synchronization. In the {\rm APD} {model}, two
chaotic parts to be synchronized can be written as the
nonautonomous form:
\begin{equation}
\dot{\textbf{z}}_1=\textbf{g}[\textbf{z}_1,\textbf{s}(t)], \qquad
\dot{\textbf{z}}_2=\textbf{g}[\textbf{z}_2,\textbf{s}(t)],
\end{equation}
where the temporal evolutions of $\textbf{z}_1$ and $\textbf{z}_2$
are ruled by the function $\textbf{g}$, and $\textbf{s}(t)$ is the
common external driving governed by the autonomous function
$\dot{\textbf{s}}(t)=\textbf{h}[\textbf{s}(t)]$. The APD model
provides a flexible method to find a proper function
$\textbf{h}[\textbf{s}(t)]$ for the complete synchronization of
chaotic systems. In this section, we use the APD model to study
the chaotic synchronization of the two optical cavity modes in an
optomechanical system.

According to the APD model, we propose two setups for realizing
the complete synchronization of chaotic optical modes in an
optomechanical system (see Fig.~\ref{fig01}). It can be seen that
both setups consist of three subsystems: (i) a strongly-driven
cavity mode, $\hat{a}_s$; (ii) two weakly-driven cavity modes,
$\hat{a}_1$ and $\hat{a}_2$; (iii) mechanical mode(s), either $b$
in setup~A, or $\hat{b}_s$, $\hat{b}_1$, and $\hat{b}_2$ in
setup~B. Here, the cavity mode $\hat{a}_s$ is strongly driven to
induce chaos. This chaos can be then transferred to the two
weakly-driven cavity modes ($\hat{a}_1$ and $\hat{a}_2$) via the
mechanical resonator(s)~\cite{OM13}. In the APD model, the
mechanical oscillation corresponds to an external signal and the
weakly-driven cavity modes are the two subsystems to be synchronized.
We show that under the common driving of the chaotic mechanical
resonator, the two weakly-driven optical modes can be excited to
chaotic states {(see Fig.~\ref{fig03})} and can evolve into a
completely-synchronized state {(see Fig.~\ref{fig04})}. For
simplicity, we neglect {both} thermal noise and quantum noise. This
is valid under the following assumptions: (i) the thermal
occupation of {the cooled} mechanical resonators is {low,} such that
the thermal noise of {the} mechanical oscillators is small in comparison
with the motion caused by the applied driving; (ii) the
optomechanical system is driven by strong laser fields and,
therefore, can be treated as a classical system. Under these
conditions, the effect of environmental thermal noise and quantum
noise of our optomechanical system can be neglected.

\subsection{Complete synchronization in setup~A}

We start our discussion of complete synchronization by introducing
setup~A, shown in Fig.~\ref{fig01}(a). One strongly and two
weakly-driven cavity modes are coupled to the same mechanical
mode. Here the strongly-driven optical mode creates mechanical
chaos through nonlinear optomechanical coupling. In this
arrangement, the fields in the weakly-driven cavity modes are
modulated in a chaotic way by the chaotic mechanical mode. The
total Hamiltonian of this synchronized system is given by (we set
$\hbar=1$ {and always assume $j=1,2$}):
\begin{eqnarray}\label{H_1A}
\hat{H}=&&\Delta_s \hat{a}_s^{\dag} \hat{a}_s + \sum_{j} \Delta_j \hat{a}_j^{\dag} \hat{a}_j + \Omega_m \hat{b}^{\dag} \hat{b} \nonumber\\
&& + i\varepsilon_s(\hat{a}_s^{\dag} - \hat{a}_s) + i\sum_{j}\varepsilon_j(\hat{a}_j^{\dag} - \hat{a}_j)\\
&& + g_s \hat{a}_s^{\dag} \hat{a}_s (\hat{b}+\hat{b}^{\dag})+
\sum_{j}g_j \hat{a}_j^{\dag}
\hat{a}_j(\hat{b}+\hat{b}^{\dag}), \nonumber
\end{eqnarray}
where $\hat{a}_s$ ($\hat{a}_j$) denotes the annihilation
operator of the strongly (weakly) driven cavity mode,
$\Delta_{s}=\omega_{{\rm cav},s}-\omega_{d,s}$
($\Delta_{j}=\omega_{{\rm cav},j}-\omega_{d,j}$) stands for the
corresponding detuning between the cavity resonance frequency
$\omega_{{\rm cav},s}$ ($\omega_{{\rm cav},j}$) and the input
laser frequency $\omega_{d,s}$ ($\omega_{d,j}$), and
$\varepsilon_s$ ($\varepsilon_j$) is the driving strength of the cavity mode $\hat{a}_s$
($\hat{a}_j$). The annihilation operator of the mechanical resonator
is represented by $\hat{b}$, and $\Omega_m$ denotes its natural
frequency. Here, $g_s$ ($g_j$) is the optomechanical single-photon
coupling strength between the cavity mode $\hat{a}_s$
($\hat{a}_j$) and the mechanical mode $\hat{b}$.

To obtain the equation of {motion of} the system in the classical
regime, we first write the quantum Langevin equations for the
Hamiltonian, given in Eq.~(\ref{H_1A}), as:
\begin{subequations}
\label{L_1A}
\begin{align}
\dot{\hat{a}}_s= & -i\Delta_s \hat{a}_s - \frac{\gamma_s}{2}
\hat{a}_s-i g_s \hat{a}_s (\hat{b}^{\dag} + \hat{b})
+\varepsilon_s \;,\\
\dot{\hat{b}}= &-i \Omega_{m}\hat{b}-\frac{\Gamma_{m}}{2}
\hat{b}-ig_s \hat{a}_s^{\dag} \hat{a}_s -ig_j \hat{a}_j^{\dag}
\hat{a}_j \;,\\
\dot{\hat{a}}_j= &-i\Delta_j \hat{a}_j - \frac{\gamma_j}{2}
\hat{a}_j - ig_j \hat{a}_j (\hat{b}^{\dag} + \hat{b}) +
\varepsilon_j\;,
\end{align}
\end{subequations}
where $\gamma_s$ ($\gamma_j$) and $\Gamma_{\!m}$ are the damping
rates of the cavity mode $\hat{a}_s$ ($\hat{a}_j$) and the
mechanical mode $\hat{b}$, respectively.

We treat the optomechanical {device} as a classical system {such that} we
can replace the quantum operators with their classical mean
values: $\alpha_s=\langle \hat{a}_s \rangle$,
$\alpha_j=\langle \hat{a}_j \rangle$, and $\beta=\langle
\hat{b} \rangle$. Note that the thermal noise and quantum noise
are neglected{,} as explained above. In this configuration,
$\alpha_j$ are the two classical cavity modes
to be synchronized, which are governed by
\begin{equation}
\begin{split}
\dot{\alpha}_j=-i\Delta_j \alpha_j - \frac{\gamma_j}{2}
\alpha_j - i G_j \alpha_j x &+ \varepsilon_j,
\end{split} \label{classical_1Aw}
\end{equation}
where $x=x_{\rm ZPF}(\beta + \beta^*)$ refers to the classical
mechanical displacement, and its nonlinear coupling strength with
the optical mode $\alpha_j$ is denoted by $G_j=g_j/x_{\rm ZPF}$.
Here $x_{\rm ZPF}$ is the zero-point fluctuation (ZPF)
displacement of the mechanical resonator.

We {apply the drivings in a way that} the radiation pressure of the
weakly-driven cavity modes $\alpha_1$ and $\alpha_2$ on the
mechanical mode $\beta$ is negligibly weak in comparison with that caused by the
strongly-driven cavity mode $\alpha_s$. In this arrangement, the
mechanical displacement $x$ is dominantly determined by the
strongly-driven optical mode. It subsequently governs the motion
of the weakly-driven cavity modes $\alpha_j$ by modulating their resonance frequencies. The
back-action from the cavity mode $\alpha_j$ on $x$ can be
neglected. {We denote this as the
unidirectional-coupling regime,} in which the
force of the weakly-driven cavity modes $\alpha_1$ and
$\alpha_2$ acting on the mechanical resonator can be
neglected. The mechanical displacement ${x}$, as an external signal, modulates these two cavity modes $\alpha_1$ and
$\alpha_2$ in the same way. As a result, the chaotic synchronization of the two
weakly-driven cavity modes $\alpha_1$ and $\alpha_2$ can be
obtained.

%

In this unidirectional-coupling regime, the motion of the cavity
mode $a_s$ and the mechanical resonator are reduced to
\begin{subequations}\label{classical_1As}
\begin{align}
\dot{\alpha}_s &=-i\Delta_s \alpha_s - \frac{\gamma_s}{2}
\alpha_s-i G_s \alpha_s x
+\varepsilon_s \;,\\
{m_{\rm eff}}\ddot{x} &= -{m_{\rm eff}} \Omega_{m}^2 x - {m_{\rm eff}} \Gamma_m \dot{x} + {\hbar G_s} |\alpha_s|^2,
\end{align}
\end{subequations}
where $m_{\rm eff}$ denotes the effective mass of the mechanical
resonator. This configuration can be understood in the APD
model as follows: The two identical cavity modes $\alpha_1$ and $\alpha_2$
are two subsystems to be synchronized. The mechanical
displacement $x$ {produces} a common external force on these two modes. These
two cavity modes are asymptotically stable. The configuration satisfies the
necessary conditions for their complete synchronization.

\subsection{Complete synchronization in setup~B}

In this subsection, we focus on setup~B shown in
Fig.~\ref{fig01}(b). Specifically, this system consists of one
strongly and two weakly-driven optomechanical systems, each of
which includes only a single cavity mode and a mechanical mode.
Different from setup~A, here the optomechanical systems are
coupled with each other via the mechanical resonators: each
mechanical mode $b_j$ in the weakly-driven optomechanical system
is coupled to the mechanical mode $b_s$ in the strongly-driven
optomechanical system with a coupling coefficient $k_j$. The total Hamiltonian of this system is described by
\begin{eqnarray}\label{H_1B}
\hat{H}=&&\Delta_s \hat{a}_s^{\dag} \hat{a}_s + \Delta_j \hat{a}_j^{\dag} \hat{a}_j + \Omega_s \hat{b}_s^{\dag} \hat{b}_s + \sum_{j}\Omega_j \hat{b}_j^{\dag} \hat{b}_j \nonumber\\
&& + g_s \hat{a}_s^{\dag} \hat{a}_s (\hat{b}+\hat{b}^{\dag})+ \sum_{j} g_j \hat{a}_j^{\dag} \hat{a}_j(\hat{b}_j+\hat{b}_j^{\dag}) \\
&& + \sum_{j}k_j(\hat{b}_j^{\dag} + \hat{b}_j)(\hat{b}_s+ \hat{b}_s^{\dag}) + i\varepsilon_s(\hat{a}_s^{\dag} - \hat{a}_s) \nonumber\\
&& + i \sum_{j} \varepsilon_j(\hat{a}_j^{\dag} -
\hat{a}_j)\;, \nonumber
\end{eqnarray}
where $\hat{a}_s$ ($\hat{a}_j$) refers to the annihilation
operator of the cavity mode in the strongly (weakly) driven
optomechanical system, $\Delta_{s}$ ($\Delta_{j}$) and
$\varepsilon_s$ ($\varepsilon_j$) are the corresponding detuning
and driving strength, {respectively}. Here, $\hat{b}_s$ ($\hat{b}_j$) is the annihilation operator of the mechanical mode in the strongly
(weakly) driven optomechanical system and its resonance frequency is
denoted as $\Omega_s$ ($\Omega_j$){;} while $g_s$ ($g_j$) is the
optomechanical single-photon coupling strength between the cavity
mode $\hat{a}_s$ ($\hat{a}_j$) and the mechanical mode
$\hat{b}_s$ ($\hat{b}_j$), and $k_j$ denotes the coupling strength
between the mechanical modes $\hat{b}_j$ and $\hat{b}_s$. Because
identical subsystems are required to achieve their complete
synchronization, the coupling coefficients are set to be the same,
$k_1=k_2$. From the Hamiltonian, given in Eq.~(\ref{H_1B}), we
have the following quantum Langevin equations:
\begin{subequations}
\label{L_1B}
\begin{align}
\dot{\hat{a}}_s= & -i\Delta_s \hat{a}_s - \frac{\gamma_s}{2}
\hat{a}_s-i g_s \hat{a}_s (\hat{b}_s^{\dag} +
\hat{b}_s)+\varepsilon_s \;,\\
\dot{\hat{b}}_s= & -i \Omega_{s}\hat{b}_s - \frac{\Gamma_{\!s}}{2}\hat{b}_s-ig_s \hat{a}_s^{\dag}
\hat{a}_s \;, \\
\dot{\hat{a}}_j= & -i\Delta_j \hat{a}_j - \frac{\gamma_j}{2} \hat{a}_j- ig_j \hat{a}_j (\hat{b}_j^{\dag}+ \hat{b}_j) + \varepsilon_j\;, \\
\dot{\hat{b}}_j= & -i \Omega_{j} \hat{b}_j - \frac{\Gamma_{\!j}}{2} \hat{b}_j-ig_j \hat{a}_j^{\dag}
\hat{a}_j  + i k_j (\hat{b}_s
+ \hat{b}_s^{\dag})\;,
\end{align}
\end{subequations}
where $\gamma_s$ ($\gamma_j$) and $\Gamma_{\!s}$ ($\Gamma_{\!j}$)
refer to the damping rates of the optical mode $\hat{a}_s$
($\hat{a}_j$) and the mechanical mode $\hat{b}_s$ ($\hat{b}_j$)
in the strongly (weakly) driven optomechanical system.
Let $\alpha_s$, $\alpha_j$, $\beta_s$, and $\beta_j$ be the
mean values of $\hat{a}_s$, $\hat{a}_j$, $\hat{b}_s$, and
$\hat{b}_j$: $\alpha_s=\langle \hat{a}_s \rangle$,
$\alpha_s=\langle \hat{b}_s \rangle$, $\beta_s=\langle \hat{b}_s
\rangle$, and $\beta_j=\langle \hat{b}_j \rangle$. In
the semiclassical regime, the operators $\hat{a}_s$,
$\hat{a}_j$, $\hat{b}_s$, and $\hat{b}_j$ in Eq.~(\ref{L_1B})
can be replaced by classical variables $\alpha_s$, $\alpha_j$,
$\beta_s$, and $\beta_j$. Here, the two classical
weakly-driven optomechanical systems to be synchronized are
governed by:
\begin{subequations}\label{classical_1Bw}
\begin{align}
\dot{\alpha}_j= &  -i\Delta_j \alpha_j - \frac{\gamma_j}{2}
\alpha_j + i G_j \alpha_j x_j +
\varepsilon_j \;,\\
{m_{\text{meff}, j}} \ddot{x}_j = & - {m_{\text{meff}, j}} \Omega_j^2 x_j - {m_{\text{meff}, j}} \Gamma_j \dot{x}_j+{\hbar G_j} |\alpha_j|^2 \nonumber \\
&- K_j(x_j-x_s)\;.
\end{align}
\end{subequations}
These two systems, described in Eq.~(\ref{classical_1Bw}), are
driven by the same strongly-driven optomechanical system:
\begin{subequations}\label{classical_1Bs}
\begin{align}
\dot{\alpha}_s= & -i\Delta_s \alpha_s - \frac{\gamma_s}{2}
\alpha_s - i G_s \alpha_s x_s + \varepsilon_s \;, \\
{m_\text{meff, s}}\ddot{x}_s = & -{m_\text{meff, s}} \Omega_{s}^2 x_s - {m_\text{meff, s}} \Gamma_s \dot{x}_s + {\hbar G_s} |\alpha_s|^2 \;,
\end{align}
\end{subequations}
where {$m_\text{meff, s}$ ($m_{\text{meff}, j}$)}, $x_s$ ($x_j$), and $G_s$ ($G_j$) are the
mechanical effective mass, mechanical displacement, and the
optical-mechanical coupling strength of the strongly (weakly)
driven optomechanical {resonator, respectively}. Here, {$x_s$ ($x_j$) is defined as}
 $x_s=x_{\rm
ZPF}^s(\beta_s + \beta_s^*)$ [$x_j=x_{\rm ZPF}^j(\beta_j +
\beta_j^*)$], and we define $G_s=g_s/x^s_{\rm ZPF}$ ($G_j=g_j/x^j_{\rm
ZPF}$), where $x_{\rm ZPF}^s$ ($x_{\rm ZPF}^j$) is the ZPF of the strongly (weakly) driven optomechanical resonators. The external force acting on the mechanical resonator {associated} with
{the} displacement $x_j$ takes the form $- K_j(x_j-x_s)$, where
$K_j = {\hbar k_j}/{(x_{\rm ZPF}^s x_{\rm ZPF}^j)}$
is the classical mechanical coupling strength. When
{$K_1/m_{\text{eff}, 1}=K_2/m_{\text{meff}, 2}$}, {the} two weakly-driven {modes} share the same
dynamics. Thus, this system can be studied in the framework of the
APD configuration: two weakly-driven optomechanical {resonators}
as two chaotic {subsystems} are synchronized and the
strongly-driven optomechanical {resonator} acts as a common external
force.

Note that the external-force term $- K_j(x_j-x_s)$ in
Eq.~(\ref{classical_1Bs}b) is derived from the classical
Lagrangian $L_{\rm int}={K_j(x_j-x_s)^2}/{2}$. If we start from
the quantum Langevin equations, {shown} in Eq.~(\ref{L_1B}d), then
the external-force term should be $K_j x_s$. This difference $-K_j
x_j$ between these two functions originates from the quantization
of classical coupled-spring oscillators. Quantum systems interact
with each other in the discontinuous regime, while the classical
ones interact in the continuous regime. When $\Omega_j^2 \gg K_j$,
the term $-K_j x_j$ in Eq.~(\ref{classical_1Bw}b) is very small
when {compared} to other terms and can be omitted.
Thus, Eqs.~(\ref{L_1B}d) and (\ref{classical_1Bw}b) are consistent
for high-frequency resonators.


\section{Phase synchronization}
\begin{center}
\begin{figure*}[t]
\centering
\includegraphics[width=6in]{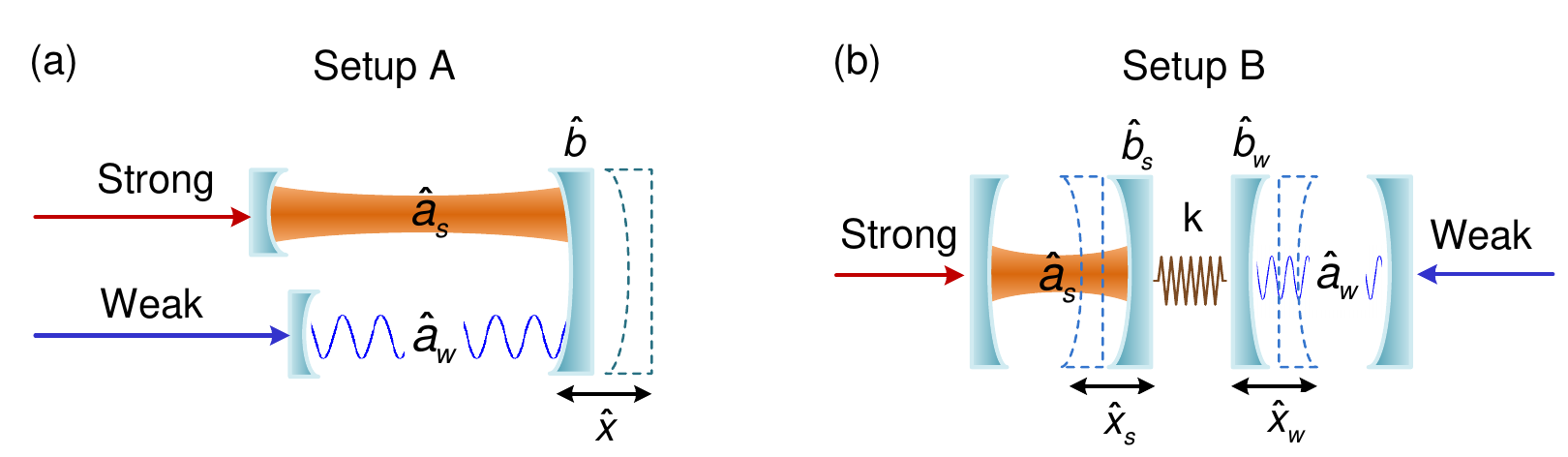}
\caption{(Color online) Schematic diagrams of two different setups
for phase synchronization. (a) Setup~A consists of a
strongly-driven optical mode {$\hat{a}_s$}, a weakly-driven optical mode {$\hat{a}_w$}, and a
mechanical mode {$\hat{b}$}. {Both} the optical modes are coupled to the
mechanical resonator and integrated into a single optomechanical
system; (b) Setup~B includes one strongly-driven {($\hat{a}_s$ and $\hat{b}_s$)} and one
weakly-driven {($\hat{a}_w$ and $\hat{b}_w$)} optomechanical systems, which are coupled via the
mechanical modes {$\hat{b}_s$ and $\hat{b}_w$} with a coupling coefficient $k$.}
\label{fig02}
\end{figure*}
\end{center}
In Sec.~\Rmnum{2}, we discussed the complete synchronization of
two identical chaotic optical modes in optomechanical systems.
However, it is technically challenging to fabricate two identical optomechanical resonators. Even a tiny parameter mismatch
between two chaotic optomechanical resonators can destroy their complete
synchronization. Thus, the research of synchronization in
nonidentical chaotic systems is {of importance}. So far, there are many attempts for synchronization of two nonidentical chaotic systems, including phase synchronization~\cite{PS1,PS2}, generalized
synchronization~\cite{GS1,GS2}, and time-delayed
synchronization~\cite{LS1}. In this section, we show that the
phases of two nonidentical chaotic cavity modes can be locked at a
fixed ratio, although their amplitudes are irrelevant to each
other.

In general, two periodic systems are called phase synchronized if
their phases $\psi_1(t)$ and $\psi_2(t)$ are locked at a fixed
ratio $m/n$, i.e.,
{$|n\psi_1(t)-m\psi_2(t)|<{\rm constant}$}, {where $m$ and $n$ are integers}. Recently, the notion of
phase synchronization has been extended to chaotic systems. {We find} that two weakly-coupled chaotic systems can be perfectly
phase locked and their amplitudes are irrelevant. {The} definition of
the phase of a chaotic system is not unique. Indeed, various
versions have been proposed based on analytic signal processing {methods}~\cite{ASP1} or {the} Poincar\'e section.
{Here, we use the former to study the phase synchronization of chaotic cavities. To obtain the temporal phase, observed in an
arbitrary-scale time function $s(t)$,} a complex analytic signal
$\phi(t)$ is reconstructed from $s(t)$, i.e.,
\begin{equation}\label{ASP}
\phi(t)=s(t)+i{\tilde{s}}(t)=A(t)\exp[i\Psi(t)],
\end{equation}
where $A(t)$ is the amplitude of the signal and {$\Psi(t)$} is its
phase, while $\tilde{s}(t)$ is the Hilbert transform of $s(t)$,
which is given by
\begin{equation}\label{}
\tilde{s}(t)=\frac{1}{\pi}{\rm
P.V.}\int^{\infty}_{-\infty}\frac{s(\tau)}{t-\tau}d\tau,
\end{equation}
where P.V. denotes the Cauchy principal value. Note that the
phases are unwrapped, i.e., these are not constrained to the range
$(-\pi,\pi]$. Once $s(t)$ is obtained, the amplitude $A(t)$ and
phase $\Psi(t)$ can be calculated~\cite{ASP1}.

In this section, we describe a configuration for phase
synchronization of a strongly-driven cavity mode
and a weakly-driven one in an optomechanical system. To do so, we propose two setups, as shown in
Figs.~\ref{fig02}(a) and~\ref{fig02}(b). Both setups
consist of one strongly-driven mode, one weakly-driven
cavity mode, and one or more mechanical mode(s). As mentioned in
Sec.~\Rmnum{2}, chaos can be generated in the strongly-driven
cavity mode and be transferred to the weakly-driven cavity modes in mediation of
the mechanical oscillation. When the driving-enhanced optomechanical coupling is strong and the cavity-driving detuning is small, {i.e.,} in the strong optical-mechanical coupling and
small-detuning regime, the temporal phase of each optical mode
mainly depends on the displacement(s) of the mechanical resonator(s). In this configuration, the two chaotic optical cavity modes can be prepared in phase synchronized, regardless of their amplitudes. As in
complete synchronization, here we also neglect thermal noise
and quantum noise.

The detailed description of these two setups
is presented in the following subsections.

%
%
%

\subsection{Phase synchronization in setup~A}

As shown in
Fig.~\ref{fig02}(a), the system consists of a strongly-driven cavity mode $\hat{a}_s$ and a
weakly-driven cavity mode $\hat{a}_w$, and a mechanical mode $\hat{b}$ {associated} with displacement $\hat{x}=x_\text{ZPF}(\hat{b}^\dag +\hat{b})$.
These two cavity modes $\hat{a}_s$ and $\hat{a}_w$ are coupled
to the mechanical mode $\hat{b}$ in the unidirectional-coupling
regime. Next we show how the phases of the two chaotic
optical modes in the strongly- and weakly-driven regimes in setup
A can be locked at a fixed ratio.

The total Hamiltonian of this system is described by
\begin{equation}\label{H_2A}
\begin{split}
\hat{H}=&\Delta_s \hat{a}_s^{\dag} \hat{a}_s + \Delta_w \hat{a}_w^{\dag} \hat{a}_w + \Omega_m \hat{b}^{\dag} \hat{b} \\
& + g_s \hat{a}_s^{\dag} \hat{a}_s (\hat{b}+\hat{b}^{\dag})+
i\varepsilon_s(\hat{a}_s^{\dag} - \hat{a}_s)\\
& + g_w \hat{a}_w^{\dag} \hat{a}_w(\hat{b}+\hat{b}^{\dag})+
i\varepsilon_w(\hat{a}_w^{\dag} - \hat{a}_w) \;,
\end{split}
\end{equation}
where $\Delta_{s}=\omega_{{\rm cav},s}-\omega_{d,s}$
($\Delta_{w}=\omega_{{\rm cav},w}-\omega_{d,w}$) is the
corresponding detuning between the cavity resonance frequency
$\omega_{{\rm cav},s}$ ($\omega_{{\rm cav},w}$) and the laser
frequency $\omega_{d,s}$ ($\omega_{d,w}$), while $\varepsilon_s$
($\varepsilon_w$) is the driving strength for the strongly
(weakly) driven cavity mode, and $\Omega_m$ is the resonance
frequency of the mechanical resonator $b$. Here, $g_s$ ($g_w$) is
the optomechanical single-photon coupling strength between the
cavity mode $\hat{a}_s$ ($\hat{a}_w$) and the mechanical mode
$\hat{b}$. For the Hamiltonian given in Eq.~(\ref{H_2A}),
we have the following quantum Langevin equations:
\begin{subequations}
\label{L_2A}
\begin{align}
\dot{\hat{a}}_s= & -i\Delta_s \hat{a}_s - \frac{\gamma_s}{2}
\hat{a}_s-i g_s \hat{a}_s (\hat{b}^{\dag} + \hat{b})
+\varepsilon_s  \label{L_2A:1}\;, \\
\dot{\hat{b}}= & -i \Omega_{m}\hat{b}-
\frac{\Gamma_{\!m}}{2}\hat{b}-ig_s \hat{a}_s^{\dag}
\hat{a}_s -ig_w \hat{a}_w^{\dag} \hat{a}_w  \label{L_2A:2} \;, \\
\dot{\hat{a}}_w= & -i\Delta_w \hat{a}_w - \frac{\gamma_w}{2}
\hat{a}_w - ig_w \hat{a}_w (\hat{b}^{\dag} + \hat{b}) +
\varepsilon_w \label{L_2A:3} \;,
\end{align}
\end{subequations}
where $\gamma_s$ ($\gamma_w$) and $\Gamma_{\!m}$ are the damping
rates of the cavity modes $\hat{a}_s$ ($\hat{a}_w$) and the
mechanical mode $\hat{b}$, respectively. Let $\alpha_s$,
$\alpha_w$, and $\beta$ be the mean values of $\hat{a}_s$,
$\hat{a}_w$, and $\hat{b}$: $\alpha_s=\langle \hat{a}_s
\rangle$, $\alpha_w=\langle \hat{a}_w \rangle$, and
$\beta=\langle \hat{b} \rangle$ in the classical regime. Their dynamics is governed by:
\begin{subequations}\label{classical_2A}
\begin{align}
\dot{\alpha}_s= & -i\Delta_s \alpha_s - \frac{\gamma_s}{2}
\alpha_s-i G_s \alpha_s x +\varepsilon_s \;, \\
\dot{\alpha}_w= & -i\Delta_w \alpha_w - \frac{\gamma_w}{2}
\alpha_w - i G_w \alpha_w x + \varepsilon_w \;,
\end{align}
\end{subequations}
where $G_s=g_s/x^s_\text{ZPF}$ ($G_w=g_w/x^w_\text{ZPF}$) represents the coupling strength between the
strongly (weakly) driven cavity mode $\alpha_s$ ($\alpha_w$)
and the mechanical mode $\hat{b}$. The cavity modes $\alpha_s$ and $\alpha_w$ are two parts to be synchronized. They are driven by the same mechanical mode $\beta$. The mechanical motion is governed by
\begin{equation}\label{classical_2Ax}
\begin{split}
{m_{\rm eff}}\ddot{x} = -{m_{\rm eff}} \Omega_{m}^2 x - {m_{\rm eff}} \Gamma_m \dot{x} + {\hbar G_s} |\alpha_s|^2,
\end{split}
\end{equation}
where $\Omega_{m}$ denotes the detuning of the mechanical mode
$\hat{b}$ and $\gamma_m$ is its damping rate.
{In our arrangement, the
effects of the weakly-driven optical modes acting on the
mechanical mode $x$ can be neglected by choosing $G_s |\alpha_s|^2\gg G_w |\alpha_w|^2$.}

{Next, we find the relation of parameters determining the ratio of the unwrapped phase of cavity modes in
phase synchronization.} We define {the} mean value of the mechanical displacement $x$ as $\bar{x}=\lim_{t\rightarrow\infty}(t-t_0)^{-1}\int_{t_0}^{t}|x(t^{\prime})| dt^{\prime}$, where $t_0$ is the initial time. We refer to the conditions $G_s \bar{x} \gg
\Delta_s$ ($G_w \bar{x} \gg \Delta_w$) and $G_s \bar{x} \gg
\gamma_s$ ($G_w \bar{x} \gg \gamma_w$) as the strong-coupling small-detuning regime. In this regime, the
instantaneous frequencies of both strong- and weakly-driven
optical modes are determined by the following {two factors: the detuning $\Delta_s$
($\Delta_w$) and the mechanical displacement-dependent parameter
$G_s x$ ($G_w x$).}  For on-resonance drivings, $\Delta_s=\Delta_w \approx 0$, the evolution of the
strongly (weakly) driven optical mode $\alpha_s$ ($\alpha_w$)
depends mainly on the mechanical motion $G_s \bar{x}$ ($G_w
\bar{x}$). Thus, the unwrapped phases, {$\Psi_w(t)$ and $\Psi_s(t)$ of the cavity modes $\alpha_s$ and $\alpha_w$, defined in Eq.~(12)
are locked at a fixed ratio of}
\begin{equation}
  \lim_{\rm t\rightarrow \infty}
  \frac{\Psi_w(t)}{\Psi_s(t)}=\frac{G_w}{G_s} \;,
 \label{lockA}
\end{equation}
{as the time approaches {infinity}.}

\subsection{Phase synchronization in setup~B}

{As shown in Fig.~\ref{fig02}(b), the setup B} consists of a strongly-driven optomechanical
system and a weakly-driven one, each of which includes only a
single cavity mode and a mechanical mode. Different from setup~A,
two optomechanical {resonators} are {mechanically} coupled with each other with a coupling coefficient
$k$. The total Hamiltonian of this synchronized system is
\begin{equation}\label{H_2B}
\begin{split}
\hat{H}=&\Delta_s \hat{a}_s^{\dag} \hat{a}_s + \Delta_w \hat{a}_w^{\dag} \hat{a}_w + \Omega_s \hat{b}_s^{\dag} \hat{b}_s + \Omega_w \hat{b}_w^{\dag} \hat{b}_w \\
& + g_s \hat{a}_s^{\dag} \hat{a}_s (\hat{b}_s+\hat{b}_s^{\dag})+ g_w \hat{a}_w^{\dag} \hat{a}_w(\hat{b}_w+\hat{b}_w^{\dag}) \\
& + k(\hat{b}_w^{\dag} + \hat{b}_w)(\hat{b}_s+ \hat{b}_s^{\dag}) + i\varepsilon_s(\hat{a}_s^{\dag} - \hat{a}_s) \nonumber\\
& + i \varepsilon_w(\hat{a}_w^{\dag} - \hat{a}_w)\;,
\end{split}
\end{equation}
where $\hat{a}_s$ ($\hat{a}_w$) denotes the annihilation
operator of the strongly-driven (weakly-driven) cavity mode,
$\Delta_{s}$ ($\Delta_{w}$) and $\varepsilon_s$ ($\varepsilon_w$)
are the corresponding detuning and driving strengths. Here,
$\hat{b}_s$ ($\hat{b}_w$) refers to the annihilation operators of
the mechanical modes in the strongly (weakly) driven
optomechanical system and $\Omega_s$ ($\Omega_w$) denotes its
resonance frequency. Each cavity mode $\hat{a}_s$ ($\hat{a}_j$)
is coupled to the mechanical mode $\hat{b}_s$ ($\hat{b}_w$) with
the coupling {strength} $g_s$ ($g_w$){, while} $k$ is the
coupling strength between mechanical modes $\hat{b}_w$ and
$\hat{b}_s$. From the Hamiltonian of Eq.~(\ref{H_2B}), we
obtain the following quantum Langevin equations:
\begin{subequations}
\label{L_2B}
\begin{align}
\dot{\hat{a}}_s= &-i\Delta_s \hat{a}_s - \frac{\gamma_s}{2}
\hat{a}_s-i g_s \hat{a}_s (\hat{b}_s^{\dag} + \hat{b}_s)
+\varepsilon_s \;, \\
\dot{\hat{b}}_s= &-i \Omega_{s}\hat{b}_s - \frac{\Gamma_{\!s}}{2}\hat{b}_s -i g_s \hat{a}_s^{\dag}
\hat{a}_s  \;,\\
\dot{\hat{a}}_w= &-i\Delta_w \hat{a}_w - \frac{\gamma_w}{2}
\hat{a}_w - ig_w \hat{a}_w (\hat{b}_w^{\dag} + \hat{b}_w) +
\varepsilon_w \;,\\
\dot{\hat{b}}_w= &-i \Omega_{w} \hat{b}_w - \frac{\Gamma_{\!w}}{2} \hat{b}_w-ig_w \hat{a}_w^{\dag}
\hat{a}_w  + i k (\hat{b}_s +
\hat{b}_s^{\dag}) \;,
\end{align}
\end{subequations}
where $\gamma_s$ ($\gamma_w$) and $\Gamma_{\!s}$ ($\Gamma_{\!w}$)
{denote} the corresponding damping rates of the optical mode
$\hat{a}_s$ ($\hat{a}_w$) and the mechanical mode
$\hat{b}_s$ ($\hat{b}_w$) in the strongly (weakly) driven
optomechanical system. Treating the whole system classically, we can replace operators with their mean values: $\alpha_s=\langle \hat{a}_s
\rangle$, $\alpha_w=\langle \hat{a}_w \rangle$,
$\beta_s=\langle \hat{b}_s \rangle$, and $\beta_w=\langle
\hat{b}_w \rangle$.  The dynamics of the
weakly-driven optomechanical {resonator} is described by:
\begin{subequations}\label{classical_2Bw}
\begin{align}
\dot{\alpha}_w= & -i\Delta_w \alpha_w - \frac{\gamma_w}{2}
\alpha_w + i G_w \alpha_w x_w +
\varepsilon_w  \;,\\
m_w \ddot{x}_w = & - {m_w} \Omega_w^2 x_w- {m_w} \Gamma_w \dot{x}_w  + {\hbar G_w} |\alpha_w|^2
 - K(x_w-x_s) \;.
\end{align}
\end{subequations}
In the unidirectional coupling regime,  the motion of the weakly-driven optomechanical {part} is governed by the strongly-driven optomechanical one. The motion of the {latter} can be modeled as
\begin{subequations}\label{classical_2Bs}
\begin{align}
\dot{\alpha}_s= & -i\Delta_s \alpha_s - \frac{\gamma_s}{2}
\alpha_s - i G_s \alpha_s x_s + \varepsilon_s \;,\\
{m_{\rm s}}\ddot{x}_s = & -{m_{\rm s}} \Omega_{s}^2 x_s - {m_{\rm
s}} \Gamma_s \dot{x}_s + {\hbar G_j} |\alpha_s|^2 \;,
\end{align}
\end{subequations}
where $G_s=g_s/x^s_{\rm ZPF}$ ($G_w=g_w/x^w_{\rm ZPF}$) is
the optical-mechanical coupling strength in the strongly (weakly)
driven optomechanical {part}. The displacements of the strongly-
and weakly-driven mechanical oscillators are given by {$x_s=x_{\rm ZPF}^s (\beta_s +
\beta_s^*)$ and $x_w=x_{\rm
ZPF}^w (\beta_w + \beta_w^*)$}, where $x_{\rm ZPF}^s$ and $x_{\rm ZPF}^w$ are the
corresponding ZPF displacements of the left and right mechanical
resonators, and $m_s$ ($m_w$) denotes the effective mass of the
mechanical resonator $b_s$ ($b_w$). Here, $- K(x_w-x_s)$ with a mechanical coupling strength $K = {\hbar k}/{(x_{\rm ZPF}^s x_{\rm ZPF}^w)}$ is the external force driving
the mechanical mode $x_w$. It provides a positive feedback to the weakly-driven mechanical mode $x_w$
when {$(x_w-x_s)<0$}. This feedback turns to be negative when {$(x_w-x_s)>0$}. Thus, when the coupling coefficient $k$
is strong enough, we have the relation: $x_w(t)\approx x_s(t)$.
We define the mean value of the mechanical displacement $x_m$ as
{$\bar{x}_m=\lim_{t\rightarrow\infty}(t-t_0)^{-1} \int_{t_0}^{t}|x_m(t^{\prime})|dt^{\prime}$},
where $t_0$ is the initial time and $m=s$ ($w$) stands for the
strongly (weakly) driven mode. In the strong-coupling small-detuning regime {when} $G_s
\bar{x}_s \gg \Delta_s$ ($G_w \bar{x}_w \gg \Delta_w$) and $G_s
\bar{x}_s \gg \gamma_s$ ($G_w \bar{x}_w \gg \gamma_w$), the
temporal phase of the strongly (weakly) driven optical mode
$\alpha_s$ ($\alpha_w$) mainly depends on $G_s \bar{x}_s$
($G_w \bar{x}_s$). {Under these approximations, the ratio of the unwrapped phases $\Psi_w(t)$ and $\Psi_s(t)$
of $\alpha_s$ and $\alpha_w$
for this setup B in the
infinite-time limit is the same as the corresponding limit, given
in Eq.~(\ref{lockA}), for setup A.}

We discuss our idea for the chaotic synchronization of optomechanical systems in the unidirectional coupling regime. This treatment is reasonable as long as $G_s |\alpha_s|^2 \gg G_s |\alpha_s|^2$. It is worth noting that both complete and phase synchronization can be obtained with slight change in the chaotic motion of the mechanical resonators when the weak force from the weakly driven cavity modes on the mechanical resonators  is taken into account.

\section{Results}

In Sec.~\Rmnum{2}, we presented four setups for both complete and
phase synchronization of chaotic optical modes in an
optomechanical system. {These} setups are different in
their configurations but share a common dynamics: the
strongly-driven cavity mode overwhelms the weakly-driven cavity
modes and drives them into chaotic motion. Now we present our
numerical results below for the configurations of these two types of
synchronization.

\subsection{Complete synchronization}

As mentioned above, we propose two setups for realizing complete
synchronization. In both setups, the two weakly-driven cavity
modes are controlled by the strongly-driven cavity mode. The
complete synchronization of our setups is described by the APD
model. The strongly-driven optical mode plays two key roles: (i) generating chaos and transferring
it to the weakly-driven optical modes, mediated by the mechanical
mode(s) and (ii) acting as a common external force on {the} weakly-driven optical modes.

\subsubsection{Complete synchronization in setup~A}
\begin{center}
\begin{figure}[t]
\centering
\includegraphics[width=\columnwidth]{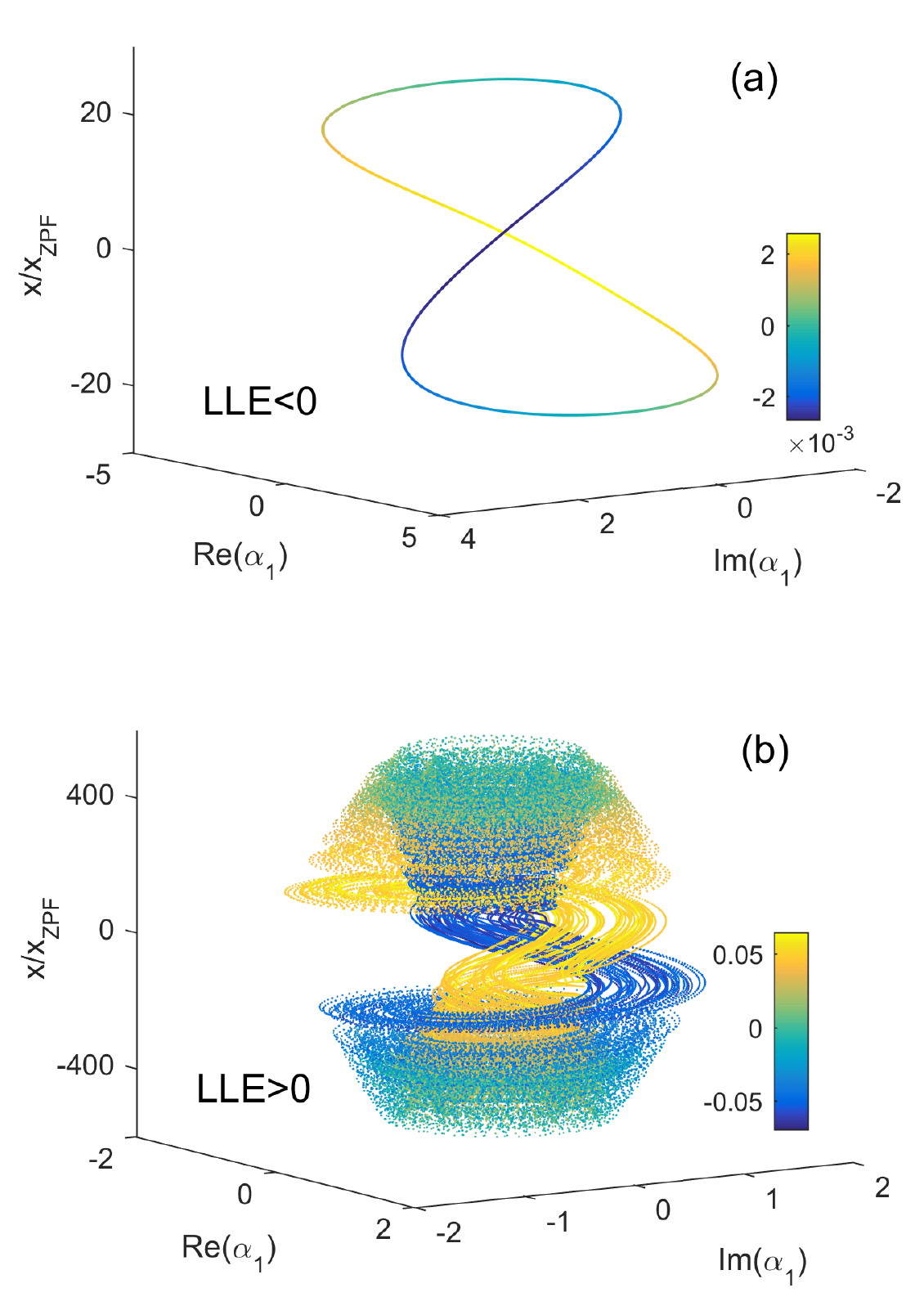}
\caption{(Color online) Complete synchronization in setup~A: The
phase portraits of optical cavity mode $\alpha_1$ and the
mechanical mode $b$ (a) without and (b) with coupling to the
strongly-driven cavity mode $\alpha_s$. The largest Lyapunov
exponent (LLE) is calculated to be negative for case (a) and
positive for case (b). Here, ${\rm Re}(\alpha_1)$, ${\rm
Im}(\alpha_1)$, and $x$ correspond to the three coordinates of
the three-dimensional phase space, and the fourth variable $p$ in
(a) and (b) is characterized in the color scale according to the
depicted colorbar. The parameters are set as:
$\Delta_1/2\pi=13~$MHz, $\gamma_1/2\pi=\gamma_s/2\pi=0.24~\rm
GHz$, $g_1/2\pi=g_s/2\pi=0.126~\rm GHz$,
$\varepsilon_1/2\pi=22~$MHz, $\Delta_s/2\pi=0.13~\rm GHz$,
$\varepsilon_s/2\pi=15.4~$GHz, $\Gamma_m/2\pi=2.8~$MHz, $m_\text{eff}=0.11~\rm fg$, and $\Omega_m/2\pi=0.346~\rm GHz$.}
\label{fig03}
\end{figure}
\end{center}

\begin{center}
\begin{figure}[t]
\centering
\includegraphics[width=\columnwidth]{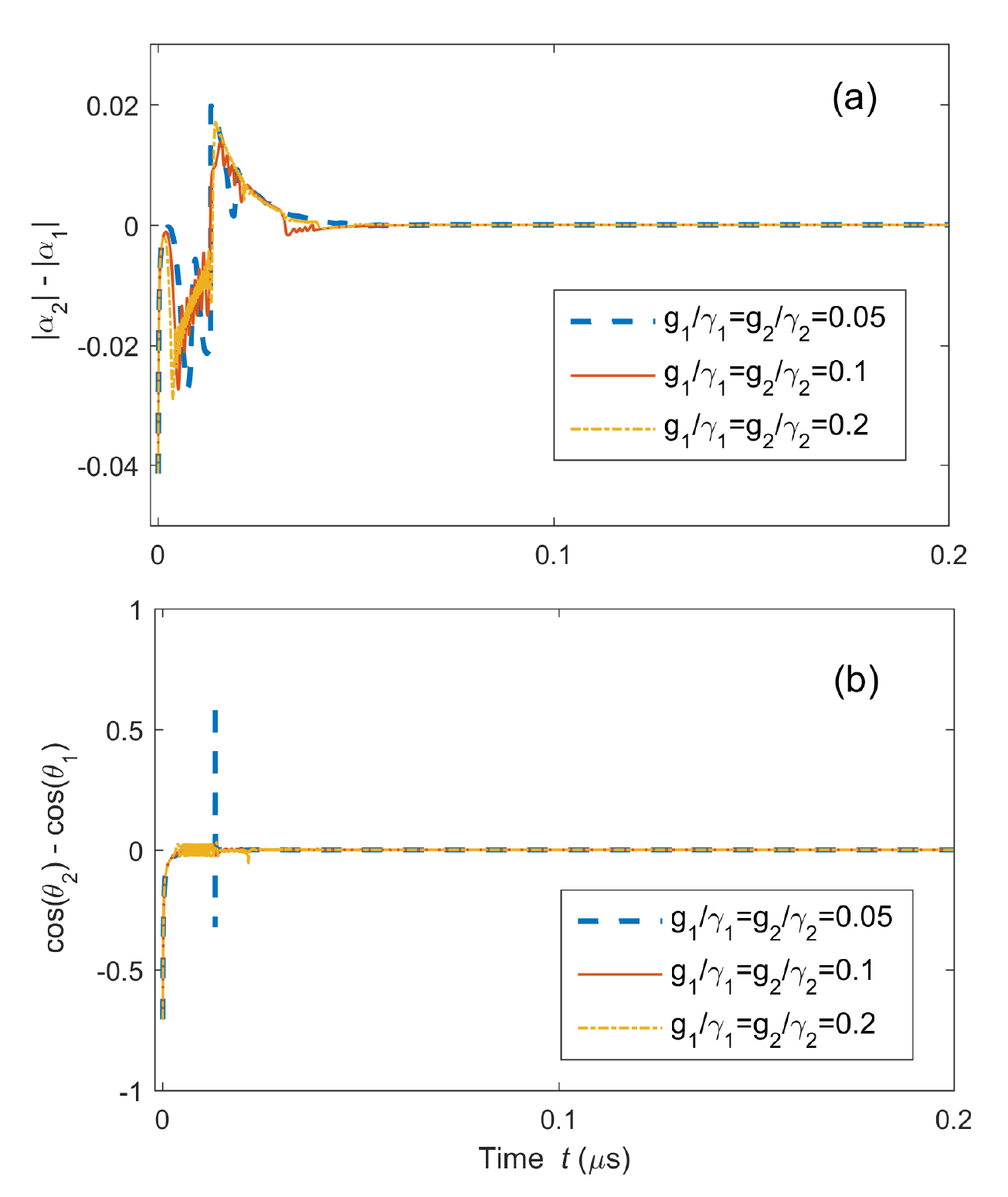}
\caption{(Color online) Synchronization errors for complete
synchronization in setup A: (a) amplitude errors and (b) phase
errors between the two chaotic weakly-driven cavity modes
$\alpha_1$ and $\alpha_2$ as a function of time $t$. Here
$\Delta_2/2\pi=13~$MHz, $\gamma_2/2\pi=0.24~\rm GHz$, and
$\varepsilon_2/2\pi=22~$MHz. The initial conditions are set as:
$\alpha_1(0) = 0.1 + 0.1 i$, $\alpha_2(0) = 0.1 i$,
$\alpha_s(0) = 0$, and $\beta(0) = 0$. All the other parameters
are the same as in Fig.~\ref{fig03}.} \label{fig04}
\end{figure}
\end{center}

In setup~A, the system consists of three cavity modes (i.e., one
strongly-driven and two weakly-driven modes) and a mechanical
mode. Each cavity mode is coupled to each other via the mechanical
mode. To realize chaotic synchronization, first, we need to
prepare the weakly-driven cavity modes $\alpha_1$ and
$\alpha_2$ in chaotic states. However, {in general,} weakly-driven
optomechanical systems can only generate nonchaotic fields. An efficient method to obtain a weak chaotic field is
that connecting a weakly-driven cavity mode to a chaotic resonator. In
this setup, chaos is generated by the strongly-driven cavity mode, and then transferred to the weakly-driven cavity modes
$\alpha_1$ and $\alpha_2$ via the mechanical mode~\cite{OM13}.
Here, the cavity mode $\alpha_1$ is taken as an example to show
how its dynamics transfers from regular into chaotic. To give a
straightforward view of this transfer, we numerically calculate its
phase portraits without [see Fig.~\ref{fig03}(a)] and with [see
Fig.~\ref{fig03}(b)] the driving from the strongly-driven optical
mode. Moreover, we calculate the largest Lyapunov exponent (LLE)
of the system with the method proposed in \cite{WOLF1985285,BRIGGS199027} to check if $\alpha_1$ evolved to a chaotic state.
A positive LLE is an indicator of chaos,
while a negative LLE means regular motion.

We first consider the case of the absence of the
cavity mode $\alpha_s$. In this case, the optomechanical system is
reduced to a single weakly-driven optical mode $\alpha_1$ and a
single mechanical mode $\beta_1$.  As shown in
Fig.~\ref{fig03}(a), a single closed loop is found in the phase
portrait with $\text{LLE}<0$, implying that the system is in regular periodic motion in the
weakly-driven regime. Then, we study the system shown in
Fig.~\ref{fig01}(a), in which the two weakly-driven cavity modes
are coupled to the strongly-driven cavity mode via the mechanical
mode. It can be seen from the phase portrait that a chaotic
attractor appears {even if} it is weakly driven [see
Fig.~\ref{fig03}(b)]. {We find that ${\rm LLE}>0$}. This means that the weakly-driven optical mode is
successfully driven to a chaotic state. The phase portraits in Fig.~\ref{fig03}(b) consist of two complex variables: the
weakly-driven cavity mode $\alpha_1$ and the mechanical mode
$\beta$. For simplicity, we expand this two-dimensional complex
space $(\alpha_1, \beta)$ to the four-dimensional real space
[{\rm Re}($\alpha_1$), {\rm Im}($\alpha_1$), $x$, $p$], where
$x$ and $p$ denote the displacement and the momentum of the
mechanical mode, respectively. The value of $p$ is presented as different colors. In Fig.~\ref{fig03}(b), we show that the weakly-driven cavity modes: (i) can
be driven to the chaotic modes and (ii) can realize the
synchronization with each other under the driving of the chaotic
mechanical resonator.


We use the synchronization error between two
chaotic fields $\alpha_1$ and $\alpha_2$ as the criterion of
complete synchronization. The synchronization error includes
the amplitude error $|\alpha_2|-|\alpha_1|$ and phase error
{$\cos \theta_2(t) - \cos \theta_1(t)$}, where $|\alpha_1|$
($|\alpha_2|$) is the amplitude of the cavity mode $\alpha_1$
($\alpha_2$), and its phase is denoted by $\theta_1(t)$
[$\theta_2(t)$]. The chaotic cavity fields $\alpha_1$ and
$\alpha_2$ are completely synchronized if both
of their amplitude and phase errors converge to zero as the evolution time
progresses to infinity. Figure~\ref{fig04} shows the synchronization error between the two chaotic
fields $\alpha_1$ and $\alpha_2$ for three different values of
the coupling strengths $g_1$ and $g_2$. Note that the initial
conditions of $\alpha_1$ and $\alpha_2$ are set to be
different. In general, two neighboring chaotic trajectories
without coupling will rapidly depart from each other because
chaos is sensitive to initial conditions. However, we can find that
both amplitude error $|\alpha_2(t)|-|\alpha_1(t)|$ [see
Fig.~\ref{fig04}(a)] and phase error {$\cos \theta_2(t)-\cos
\theta_1(t)$} [See Fig.~\ref{fig04}(b)] decrease to zero after
conquering the transient states. Thus, complete
synchronization is obtained in the two weakly-driven cavity
modes and this synchronization is independent of the coupling
strengths $g_1$ and $g_2$. Note that if the coupling strength
$g/\gamma_1<0.05${,} then the weakly-driven cavity mode $\alpha_1$
cannot be driven to a chaotic state.

\subsubsection{Complete synchronization in setup~B}

\begin{center}
\begin{figure}[]
\centering
\includegraphics[width=\columnwidth]{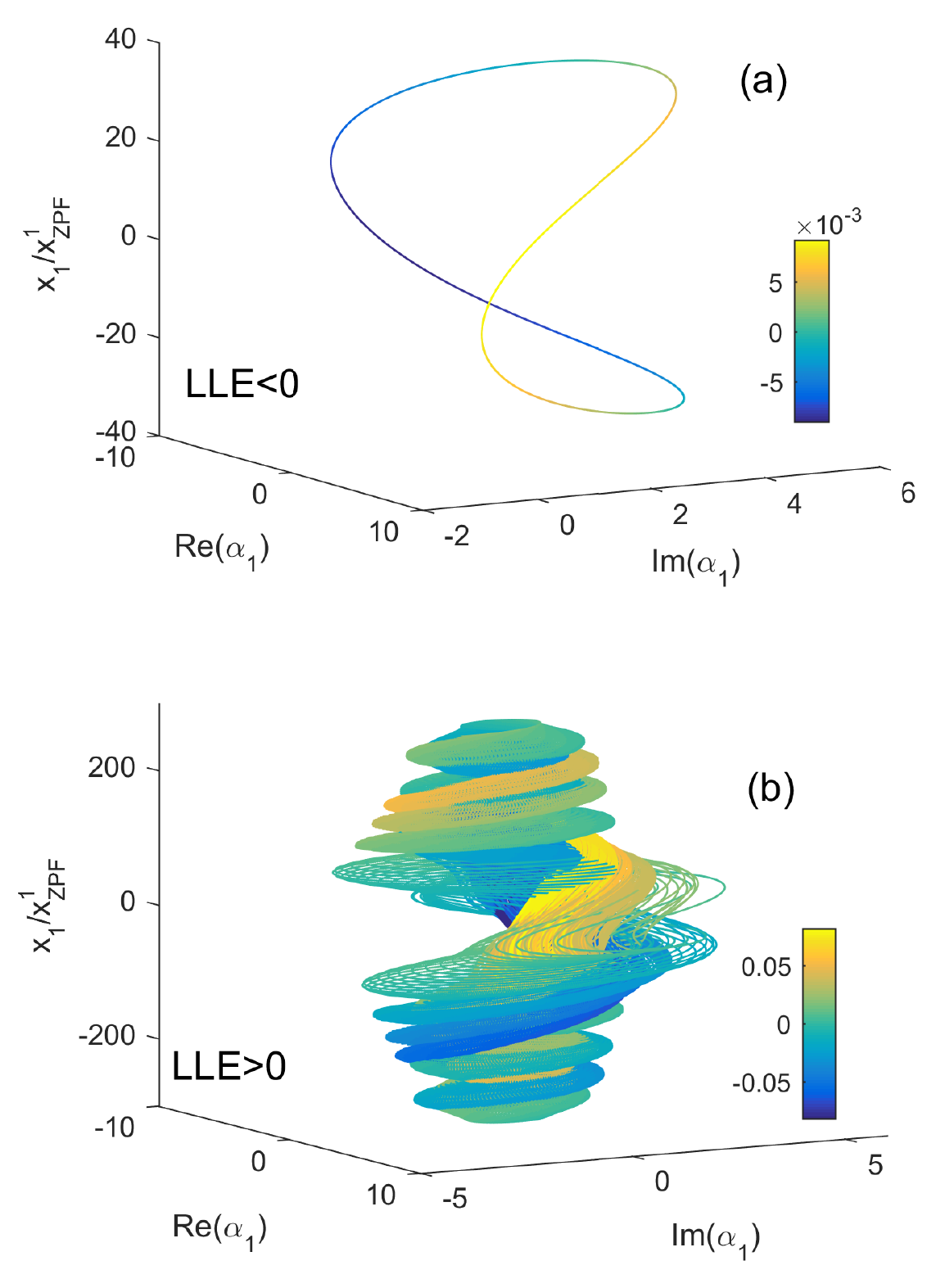}
\caption{(Color online) Complete synchronization in setup B: The
phase portraits of an optomechanical system includes cavity~1 and
the mechanical mode (a) without and (b) with coupling to the
strongly-driven cavity mode. The largest Lyapunov exponent is
calculated as: (a) ${\rm LLE}<0$ and (b) ${\rm LLE}>0$. Here ${\rm
Re}(\alpha_1)$, ${\rm Im}(\alpha_1)$, and $x_1$ correspond to
the three coordinates of the three-dimensional phase space, and
the fourth variable $p_1$ in (a) and (b) are characterized by
different colors shown according to the colorbars. {The parameters
are: $\Delta_1/2\pi=26~$MHz, $g_1/2\pi=25.2~$MHz, $\Gamma_s/2\pi=\Gamma_1/2\pi=2.8~$MHz,
$\Omega_s/2\pi=\Omega_1/2\pi=0.346~\rm GHz$, and $k_1/2\pi=k_2/2\pi=1.29~\rm
MHz$, while other parameters {are the same} as in Fig.~3.}} \label{fig05}
\end{figure}
\end{center}
Now {we study} complete synchronization in setup~B{,} shown in
Fig.~\ref{fig01}(b). The system includes three optomechanical
{subsystems}: two weakly-driven optomechanical {objects} are coupled to
the strongly-driven optomechanical one via the mechanical
coupling. Here, the two optical modes $\alpha_1$ and
$\alpha_2$ in two weakly-driven {parts} are chaotic and will
be synchronized. In this subsection, we numerically show, by
preparing the strongly-driven optomechanical {part} in a chaotic state, that the
weakly-driven parts can also be driven into synchronized chaotic states.

Since the two weakly-driven {components} share the same dynamics,
we choose one of them as an example to show the transfer from regular into chaotic motion.  The phase portraits and the LLE of the weakly-driven optomechanical
{part} $(\alpha_1,\beta_1)$ are calculated {for the cases} without and with the
strongly-driven optomechanical {resonator}. {First, in the former case, }a single loop is seen in the phase portrait {shown in Fig.~\ref{fig05}(a)}. {This implies that the
weakly-driven optomechanical part is in a periodic motion.} This
single loop becomes a chaotic attractor [see
Fig.~\ref{fig05}(b)] when the two weakly-driven optomechanical {parts} are coupled to the strongly-driven one. {Moreover, this transition is also indicated by LLE, changing from negative in Fig.~\ref{fig05}(a) to positive in  Fig.~\ref{fig05}(b).} Similarly to Fig.~\ref{fig03}, the complex two-dimensional
weakly-driven optomechanical {resonator} [$\alpha_1$, $\beta_1$] is
illustrated in the four-dimensional real space [{\rm
Re}($\alpha_1$), {\rm Im}($\alpha_1$), $x_1$, $p_1$]. Here,
{\rm Re}($\alpha_1$) [{\rm Im}($\alpha_1$)] is the real
(imaginary) part of the classical cavity mode $\alpha_1$, and
$x_1$ ($p_1$) denotes the displacement (momentum) of the classical
mechanical mode $\beta_1$. {The color of the lines show the values of the fourth component $p_1$.}
\begin{center}
\begin{figure}[]
\centering
\includegraphics[width=\columnwidth]{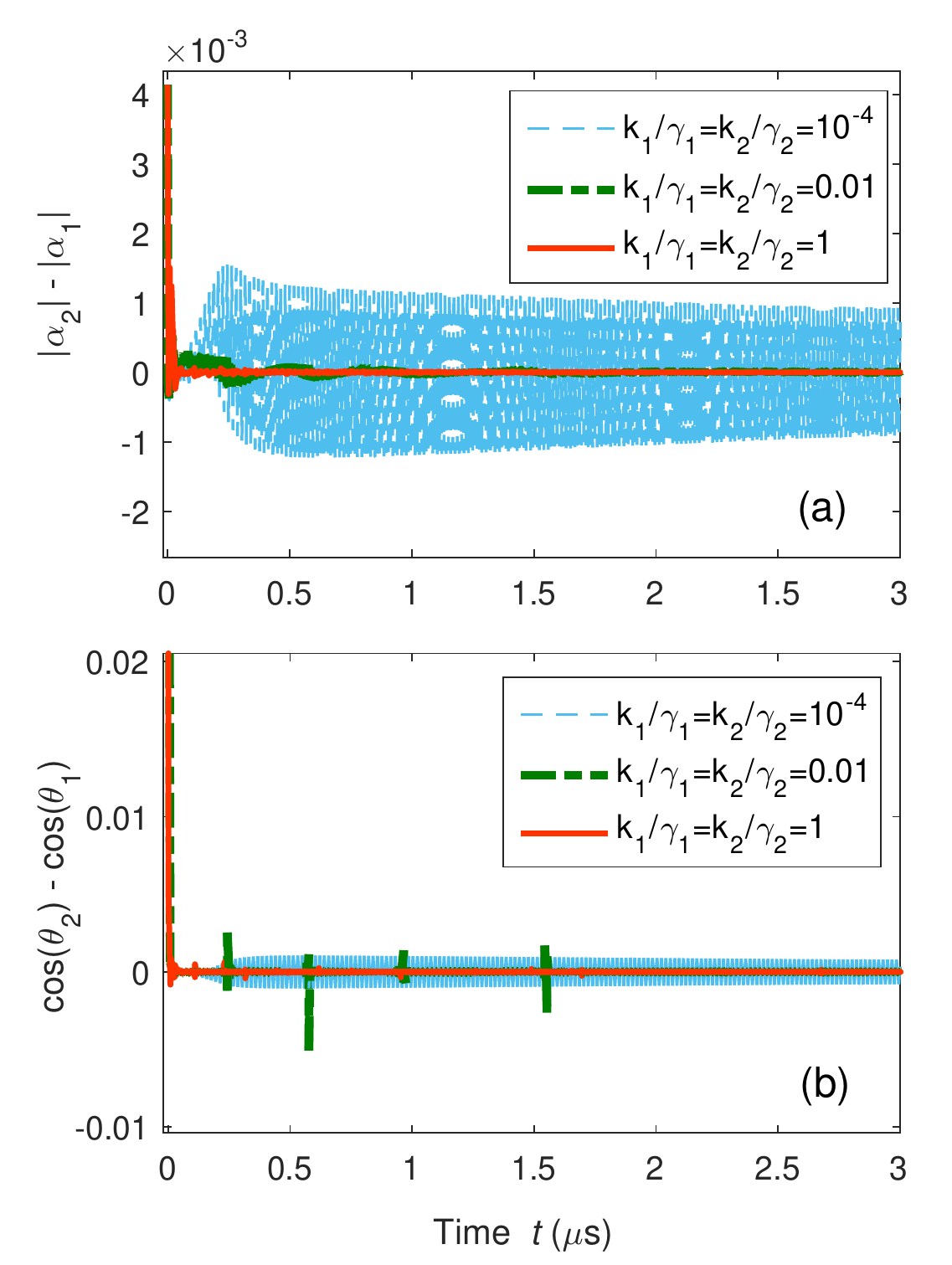}
\caption{(Color online) Synchronization errors for complete
synchronization in setup B: (a) amplitude errors and (b) phase
errors between the cavity modes $\alpha_1$ and $\alpha_2$ for
different mechanical-mechanical coupling coefficients $k_1$ and
$k_2$ as a function of time $t$. The initial conditions of the
weakly- and strongly-driven optomechanical systems are set as:
$[\alpha_1(0), \beta_1(0)] = (0.01 i, 0)$, $[\alpha_2(0),
\beta_2(0)] = (0.01 + 0.01 i, 0)$, and $[\alpha_s(0),
\beta_s(0)]= (0, 0)$. The other parameters are the {same as in}
Fig.~\ref{fig05}. } \label{fig06}
\end{figure}
\end{center}


To answer the question whether the two chaotic cavity modes
$\alpha_1(t)$ and $\alpha_2(t)$ can achieve complete synchronization, we calculate their error and check if it
converges to zero. The error here includes the
amplitude error $|\alpha_2|-|\alpha_1|$ and the phase error
$\cos \theta_2(t)-\cos \theta_1(t)$, where $|\alpha_1|$
($|\alpha_2|$) and $\cos \theta_1(t)$ [$\cos \theta_2(t)$]
denote the amplitude and phase of the cavity mode $\alpha_1$
($\alpha_2$), respectively.
Figures~\ref{fig06}(a) and~\ref{fig06}(b) show the amplitude
and phase errors for different mechanical-coupling coefficients
$k_1$ and $k_2$. The initial condition difference is set to be:
$|\alpha_2|-|\alpha_1|=0.0041$ in Fig.~\ref{fig06}(a) and
$\cos \theta_2(t)-\cos \theta_1(t)=1/\sqrt{2}$ in
Fig.~\ref{fig06}(b). When $k_1$ and $k_2$ are very weak
($k_1/\gamma_1=k_2/\gamma_2=10^{-4}$), both amplitude
and phase errors considerably fluctuate [blue dashed curves in
Figs.~\ref{fig06}(a) and~\ref{fig06}(b)] as the evolution time
progresses. When $k_1$ ($k_2$) increases to
$k_1/\gamma_1=k_2/\gamma_2=10^{-2}$ [green dashed dot curves in
Fig.~\ref{fig06}(a) and~\ref{fig06}(b)], these two
errors drastically fluctuate in the beginning, and then decrease
to zero after conquering a transient period. Moreover, the increase of the coupling strength $k_1$ ($k_2$)
accelerates the convergence of the synchronization errors, as
shown in Figs.~\ref{fig06}(a) and~\ref{fig06}(b) (red solid
curve). The time going to synchronization greatly decreases as
the parameters $k_1$ and $k_2$ increase to $k_1/\gamma_1=k_2/\gamma_2=1$ from $k_1/\gamma_1=k_2/\gamma_2=10^{-2}$. {Obviously}, the mechanical-coupling parameters $k_1$ and $k_2$
play a crucial role in the synchronization of chaotic optical
fields. Two weakly-driven optomechanical systems can be driven into complete synchronization when the mechanical coupling $k_1$ and $k_2$ are large enough.

\subsubsection{Comparison of setup~A and setup~B}

Complete synchronization can be realized in both setups~A and
B according to the APD model. As shown in
Figs.~\ref{fig04} and~\ref{fig06}, the motions of the two
weakly-driven cavity modes tend to be close to each other and
become completely identical as the time progresses. As our theoretical prediction,  the weakly-driven
cavity modes in chaotic motion can be {in complete synchronization} if they are
asymptotically stable and their motion is dominated by a common
external force, which is the strongly-driven cavity mode here.
Chaos can be transferred from the strongly-driven cavity mode to the weakly-driven cavity modes
by mediation of a direct coupling in setup~A (see Fig.~\ref{fig03}) or
indirect coupling in setup~B (see Fig.~\ref{fig05}).
In setup~A, the two weakly-driven cavity modes
are synchronized. They are driven by the same
mechanical mode.

Different from setup~A, the action of the common
external drive in setup~B is indirectly applied to the two weakly-driven cavity modes via the mechanical coupling.
The setup B highly relies on the mechanical coupling coefficient
$k_1$ and $k_2$. The motion of the optical modes of
weakly-driven optomechanical systems is not only affected by its own oscillation but more crucially depends on the
strongly-driven optomechanical {one}. When the
mechanical-coupling coefficients are large,
complete synchronization is achieved.
However, for small $k_1$ and $k_2$, the motion of the mechanical
modes is dominated by the weakly driven optical modes. Thus, the external drive has little
effect on the optical cavity modes to be synchronized. As a result, in the
weak mechanical-coupling regime, complete synchronization is impossible in setup B.

\subsection{Phase synchronization}

Phase synchronization is defined as the locking of the unwrapped
phases in two dynamical systems. Below we will show phase synchronization of two chaotic optical modes in Fig.~\ref{fig02} in the strong-coupling small-detuning regime.
Note that the unwrapped phases defined here are unfolded in every
$2\pi$-period. This is essentially different from the phases
introduced for complete synchronization in Sec.~\Rmnum{2}.

\subsubsection{Phase synchronization in setup~A}
\begin{center}
\begin{figure}[t]
\centering
\includegraphics[width=\columnwidth]{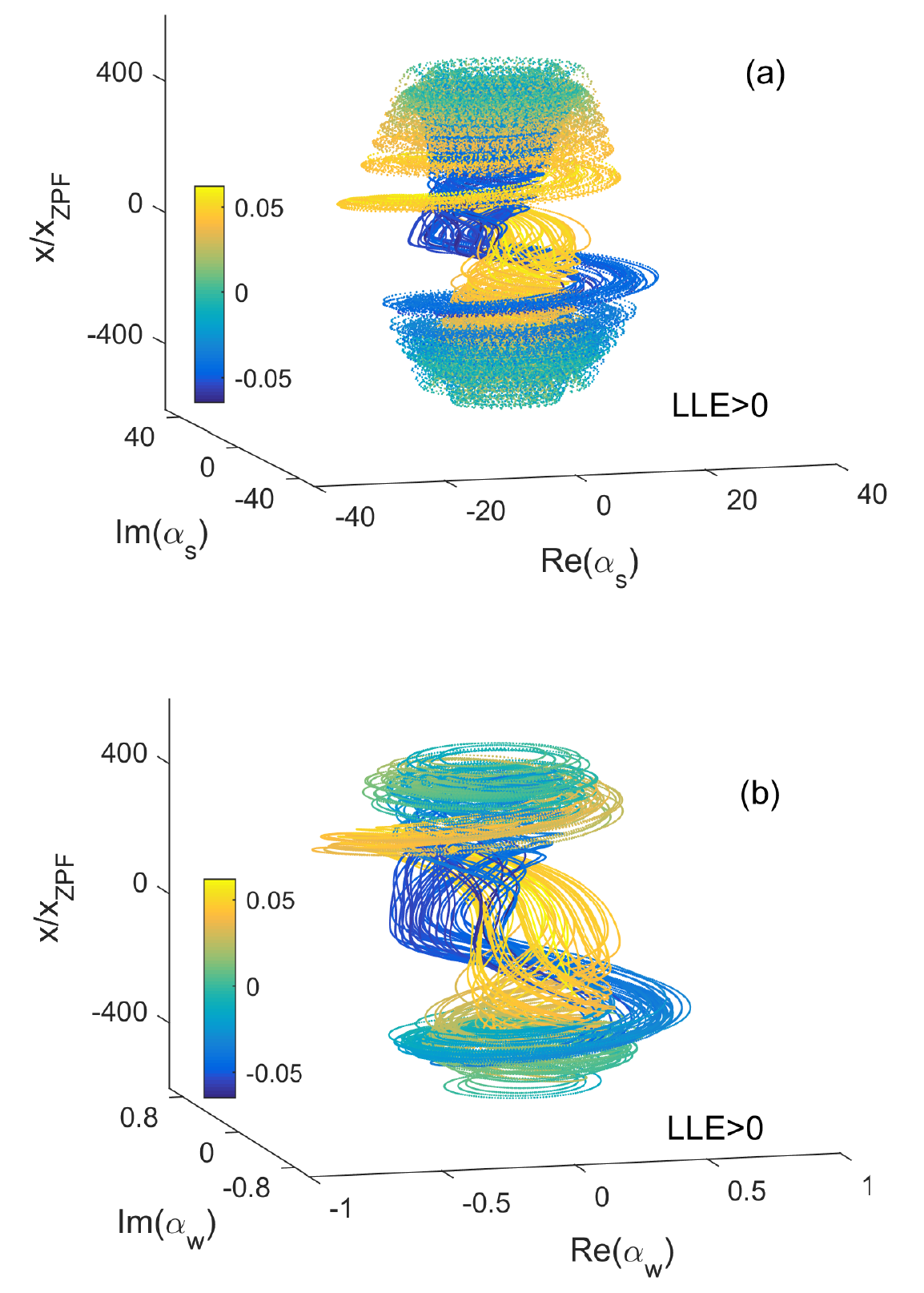}
\caption{(Color online) Phase synchronization in setup A:  The
phase portraits of (a) the strongly-driven and (b) weakly-driven
optomechanical system. {The largest Lyapunov exponent is positive in both cases.} The parameters here
are: $\Delta_s/2\pi=0.13~\rm GHz$, $\gamma_s/2\pi=0.24~\rm GHz$,
$g_s/2\pi=0.126~\rm GHz$, $\varepsilon_s/2\pi=15.4~\rm GHz$,
$\Delta_w/2\pi=26~$MHz, $\gamma_w/2\pi=52~$MHz, $g_w/2\pi=25.2~\rm
MHz$, $\varepsilon_w/2\pi=0.22~\rm GHz$, $\Gamma_m/2\pi=2.8~\rm
MHz$, and $\Omega_m/2\pi=0.346~\rm GHz$.} \label{fig07}
\end{figure}
\end{center}

\begin{center}
\begin{figure}[t]
\centering
\includegraphics[width=\columnwidth]{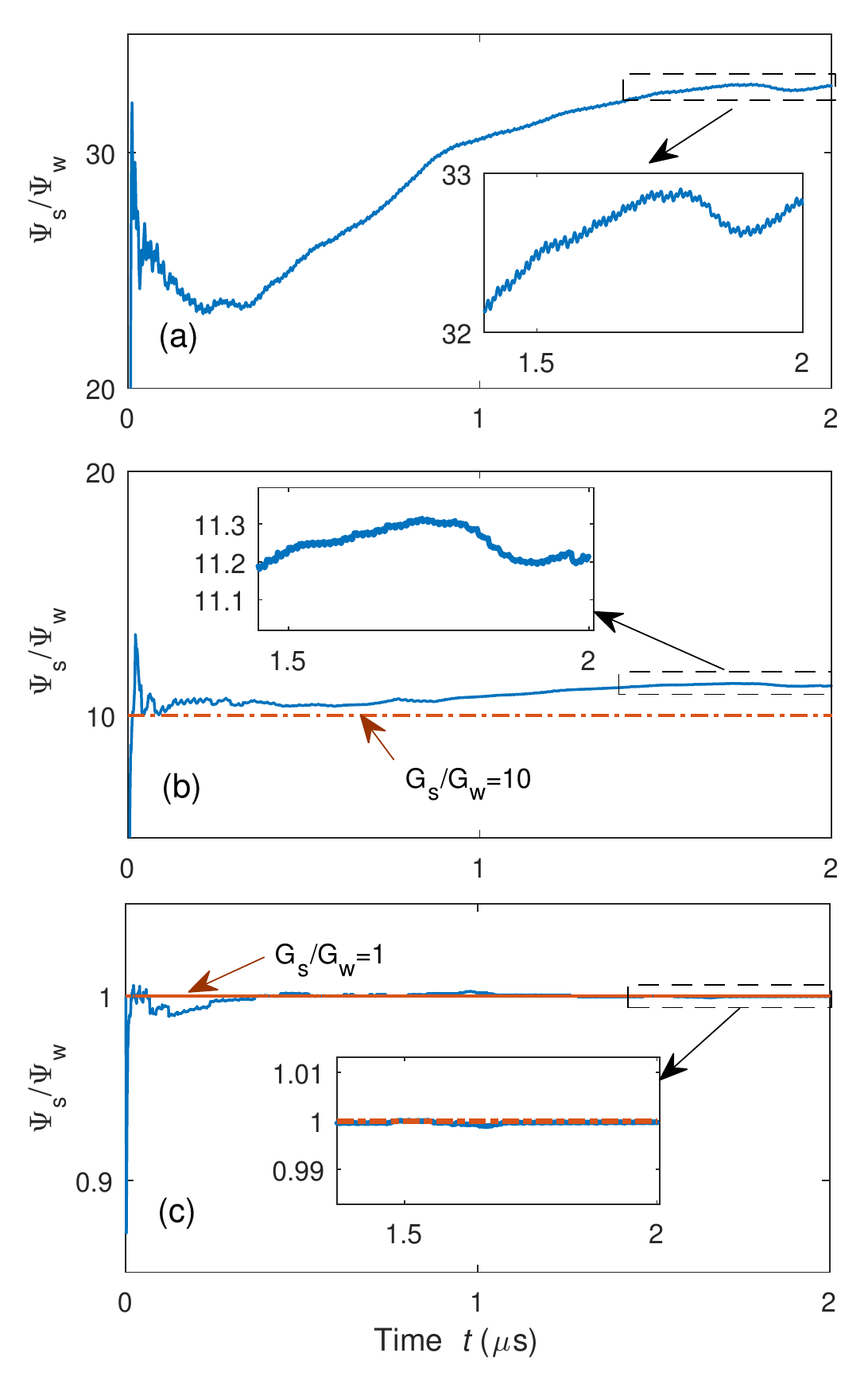}
\caption{(Color online) Phase synchronization in setup A:
Evolutions of the ratios for the phases of the strongly-
[$\Psi_s(t)$] and weakly-driven $\Psi_w(t)$ cavity modes, when
their coupling {strengths} are (a) {$G_s/G_w=100$, (b)
$G_s/G_w=10$, and (c) $G_s/G_w=1$}, where $G_s=g_s/x_{\rm ZPF}$ is
a fixed value and $G_w=g_w/x_{\rm ZPF}$. The red dashed line in
each panel denotes the forecasting value $G_s/G_w$. Here,
$\varepsilon_w/2\pi=1.1~\rm GHz$, $g_s=0.126~\rm GHz$, and the
coupling {strengths} between the weakly-driven optical mode and the
mechanical resonator are {(a) $g_w/2\pi=1.26~$MHz, (b) $g_w/2\pi=12.6~$MHz,
and (c) $g_w/2\pi=0.126~\rm GHz$.} The other parameters are the same as
in Fig.~\ref{fig07}.} \label{fig08}
\end{figure}
\end{center}

In setup A{,} shown in Fig.~\ref{fig02}(a), the weakly- and strongly-driven
optical modes are coupled via a mechanical mode. When the cavity mode $\hat{a}_s$ is strongly driven into a chaotic state, it, in turn, brings the mechanical mode into chaotic motion. As a result, the weakly-driven
cavity mode is driven to a chaotic state via its coupling
to the mechanical mode. The chaotic motion of two cavity modes is also proved by the positive LLE. In spite of  being in chaotic motion, the motion trajectories of two chaotic optical {modes} have dramatically different amplitudes. However, two attractors rotate in a similar way with respect to the axis of $\alpha_s=0$ in Fig.~\ref{fig07}(a) and $\alpha_s=0$ in Fig.~\ref{fig07}(b), respectively. It indicates {a} correlation {of the phases in the two attractors.}

To study phase synchronization between the two chaotic
optical modes, we calculate the ratio of the unwrapped phases of the
strongly- and weakly-driven optical modes.
To do so, we fix $G_s$ but change
the coupling strength $G_w$ to see how the optomechanical coupling
strength influences phase synchronization in the optomechanical
system. The unwrapped phase $\Psi_w(t)$ [$\Psi_s(t)$] of the
weakly (strongly) driven cavity mode is evaluated from the real
part of the observed signal ${\rm Re}[\alpha_w(t)]$ (${\rm
Re}[\alpha_s(t)]$) with the {analytic signal processing method.} Phase synchronization
occurs if the ratio of the phases of two nonidentical optical
modes can be locked at a fixed value of $G_s/G_w${,} as $t
\rightarrow \infty${,} according to our discussion in Sec.~\Rmnum{2}.

Figure~\ref{fig08} illustrates the evolutions of the ratio of unwrapped phase
$\Psi_s(t)/\Psi_w(t)$ as a function of the coupling strength $G_w$. When
$G_w$ is very weak, e.g. {$G_s/G_w=100$}, the motion of the
weakly-driven optical mode mainly depends on a given periodic
input field. As a result, {the ratio of $\Psi_s(t)/\Psi_w(t)$ fluctuates over a large
region $[32,34]$ and does not converge, see Fig.~\ref{fig08}(a).}
When $G_w$ is larger (e.g. $G_s/G_w=10$), [See
Fig.~\ref{fig08}(b)], the ratio of $\Psi_s(t)/\Psi_w(t)$ fluctuates within a relative smaller region, but still cannot {approach} to a constant value [see Fig.~\ref{fig08}(b)] because the {influence} of the input field and the driving
of the mechanical mode on the weakly-driven optical
mode compete with each other,
leading to the randomly varying rhythms of the strongly- and
weakly-driven cavity modes. In the
strong-coupling regime, e.g. $G_s/G_w=1$, the phase of the weakly-driven cavity mode
is dominantly controlled by the chaotic mechanical mode. This mechanical mode also
acts on the strongly-driven cavity mode simultaneously.
In this case, the resonance frequencies of both the weakly- and strongly-driven cavity
modes are determined by the motion of the mechanical mode, see
Fig.~\ref{fig08}(c). The phase ratio converges to a constant value after oscillating over a transient period. These results show that phase synchronization can be realized in two chaotic optical
oscillators, whereas their amplitudes are quite different.
Moreover, the fixed value here approximately equals to the ratio of
optomechanical strengths $G_s/G_w=1$ [red dashed line in
Fig.~\ref{fig08}(c)], {consistent} with our theoretical
analysis.

\subsubsection{Phase synchronization in setup~B}

In setup~B shown in {Fig.~\ref{fig02}(b),} the strongly- and weakly-driven optomechanical
systems are coupled to each other with a rate $k$ via the
mechanical coupling between two mechanical oscillators. When $k$ is strong enough, the motion of
the weakly-driven optomechanical {component} (right-hand optomechanical {resonator}) is dominantly controlled by the
strongly-driven optomechanical {component} (left-hand side one). Below, we will show, in the strong-coupling, small-detuning regime, the phases of the strongly- and weakly-driven cavity modes can be locked at a fixed ratio.

\begin{center}
\begin{figure}[t]
\centering
\includegraphics[width=\columnwidth]{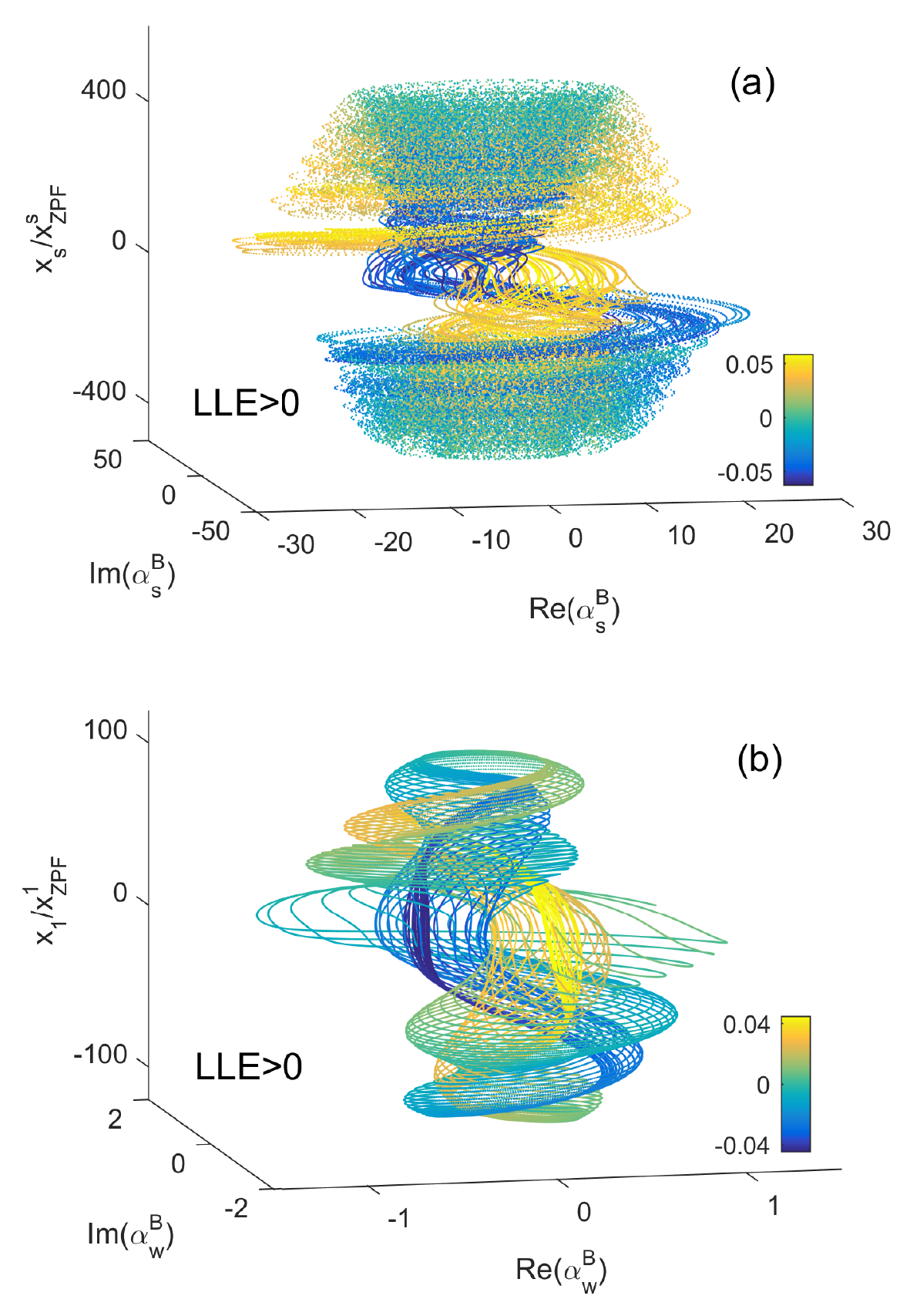}
\caption{(Color online) Phase synchronization in setup B: Phase
portraits of (a) the strongly-driven and (b) the weakly-driven
optomechanical systems. {The largest Lyapunov exponent is positive in both cases.} The parameters here are:
$\Gamma_s/2\pi=\Gamma_w/2\pi=2.8~$MHz, $\Omega_s/2\pi=\Omega_w/2\pi=0.346~\rm GHz$, and $k/2\pi=1.29~$MHz, while other parameters are the same as in Fig.~\ref{fig07}.}
\label{fig09}
\end{figure}
\end{center}
\begin{center}
\begin{figure}[t]
\centering
\includegraphics[width=\columnwidth]{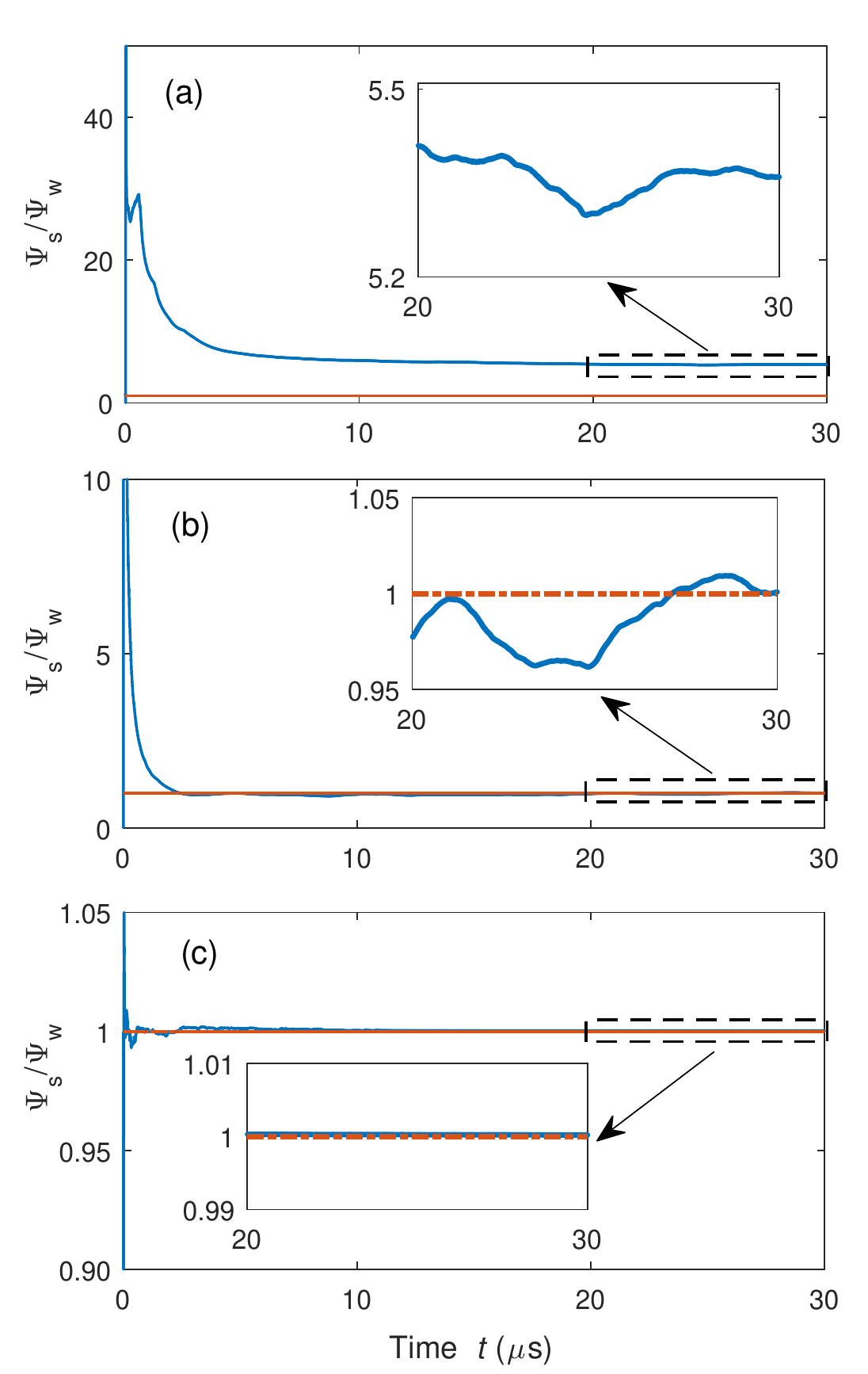}
\caption{(Color online) Phase synchronization in setup B: The
ratios of the phases between the strongly- and weakly-driven
optomechanical systems by varying the mechanical-coupling coefficient $k$: {(a) $k/\gamma_s=10^{-3}$, (b)
$k/\gamma_s=10^{-2}$, and (c) $k/\gamma_s=10^3$. Here, $g_s/2\pi=g_w/2\pi=0.126~\rm GHz$. All the other
parameters are the same as in Fig.~\ref{fig09}.} The red line
denotes the ratio of $G_s/G_w$.}
\label{fig10}
\end{figure}
\end{center}

As an example, we take the set {of} values for parameters in Fig.~\ref{fig09} for {the} numerical simulation of {the} system motion with Eqs.~(22) and (23).
As shown in Fig.~\ref{fig09}(a), the phase portrait of the strongly-driven cavity mode shows a typical chaotic attractor and its
LLE is positive. This chaotic cavity mode drives the mechanical oscillator, ${x}_s$, into chaotic motion. Due to the strong mechanical coupling between ${x}_s$ and ${x}_w$, the mechanical oscillator ${x}_w$ and subsequently the associated cavity mode under a weak driving are brought into chaotic motion. The chaotic motion of the
weakly-driven optomechanical {part} can be seen in Fig.~\ref{fig09}(b). The corresponding  LLE is also positive, as an indicator of chaotic attractor.

To check if phase synchronization can be realized in setup~B, we
numerically calculate the time evolution of the unwrapped phases of
the two cavity modes in the strongly- and weakly-driven
optomechanical systems. {Again,} with the {analytic signal processing method}, we calculate the phases $\Psi_w(t)$ and $\Psi_s(t)$ from the
optical signals ${\rm Re}[\alpha_w(t)]$ and ${\rm
Re}[\alpha_s(t)]$ by Eq.~(\ref{ASP}). Basically, the motion of
the weakly-driven optomechanical {part} is determined by two
factors: (i) its inherent oscillation and (ii) the driving of the
strongly-driven optomechanical {resonator}. For the latter factor, the
mechanical coupling coefficient $k$ acts as the coupling strength
between the strongly- and weakly-driven {components}. Here, we focus on
the influence of $k$ on phase synchronization.

Figure~\ref{fig10} shows the temporal evolution of the phase ratio, $\Psi_w(t)/\Psi_s(t)$, of two
optical cavity modes for different mechanical-coupling strengths
$k$. As mentioned above, in phase synchronization, the unwrapped phases of the two chaotic
optical cavity modes should be locked at the value $G_s/G_w$, which refers to the coupling strength of the
strongly-driven optomechanical {part}. To study the influence of the mechanical coupling on the phase synchronization, we set $G_s/G_w=1$ here.
When $k$ is very small ($k/\gamma_s=10^{-3}$), see Fig.~\ref{fig10}(a), the motion of the
weakly-driven optomechanical subsystem is separable from the
strongly-driven one. As a result, its motion is mainly determined by itself.
Thus, the phases of the two cavity modes in two {parts} are
uncorrelated. {The phase ratio
$\Psi_s(t)/\Psi_w(t)$ fluctuates in the range
$[5.2,5.5]$ as the evolution time increases.} As
$k/\gamma_s$ increases to $10^{-2}$, the ratio
$\Psi_s(t)/\Psi_w(t)$ oscillates around but cannot stay at the
value $G_s/G_w=1$ as the evolution time progresses [see
Fig.~\ref{fig10}(b)]. In this case, the phase of the
weakly-driven optomechanical system is mainly dependent on its own oscillation and the external driving
force. It can be seen in the inset of Fig.~\ref{fig10}(b) that
$\Psi_s(t)/\Psi_w(t)$ fluctuates in a much smaller range
$[0.95,1.05]$, compared to the case in Fig.~\ref{fig10}(a). When $k/\gamma_s=10^3$, the motion of the weakly-driven cavity mode is
governed by the strongly-driven optomechanical system. It leads to
a perfect phase locking, as shown in Fig.~\ref{fig10}(c). Note
that there still exists a small discrepancy between
$\Psi_s(t)/\Psi_w(t)$ and $G_s/G_w$, mainly because
the temporal phases of the optical cavity modes are also effected
by its own oscillation. This phase mismatch decreases as the
mechanical coupling coefficient $k$ increases.

\subsubsection{Comparison of setups A and B}

Both setups A and B can be described as a common configuration in
which the strongly-driven optical mode dominates the motion of the
weakly-driven optical mode. To realize phase synchronization,
setup~A requires strong optomechanical coupling and weak detuning
(the so-called strong-coupling small-detuning regime). Compared to
setup A, the setup B additionally requires a strong coupling between the two
mechanical resonators.

\section{Conclusions and discussions}

We {have} studied both complete and phase synchronization of optical
cavity modes mediated by mechanical resonators. It is
found that the complete synchronization of two identical optical
cavity modes in chaotic motion can be obtained. We also
showed the phase synchronization between two nonidentical
optomechanical systems. In both types of chaotic synchronization,
the chaotic displacement of the mechanical resonators is
dominantly governed by the strongly-driven optical mode. The chaotic
motion of the mechanical resonators subsequently pulls the
weakly-driven optical cavity modes into chaotic motion. As a
result, the phases of the strongly- and weakly-driven cavity modes can be
synchronized. Our work provides a method to observe chaotic
synchronization in experimentally-accessible optomechanical
systems.

\section*{acknowledgments}
The authors thank Yu-Xi Liu, Jing Zhang, Xuedong Hu, and Wei Qin
for useful discussions. KX would like to thank the National Key
R\&D Program of China (Grant No. 2017YFA0303703). AM and FN {are}
partially supported by the MURI Center for Dynamic Magneto-Optics
via the AFOSR Award No. FA9550-14-1-0040, the Japan Society for
the Promotion of Science (KAKENHI), the IMPACT program of JST,
CREST Grant No. JPMJCR1676,  RIKEN-AIST Challenge Research Fund,
JSPS-RFBR Grant No. 17-52-50023, and the Sir John Templeton
Foundation.


%

\end{document}